\documentclass[11pt]{article}
\usepackage{rotating}
\usepackage{comment}
\usepackage{longtable}
\usepackage{lscape}
\usepackage{amsfonts}
\usepackage{amsmath}
\usepackage{pdfpages}
\usepackage{booktabs,caption}
\usepackage[flushleft]{threeparttable}
\usepackage[utf8]{inputenc}
\usepackage{graphicx}
\usepackage{lscape}
\usepackage{url}
\usepackage{ragged2e}
\usepackage{subfig}

\title{Gentrification, Mobility, and Consumption}

\author{Giacomo De Giorgi\footnote{Giacomo De Giorgi acknowledges financial support from the SNF grant $\#100018\_182243$.  We thank Yujung Hwang, Pietro Campa, Davide Pietrobon, Jaromir Nosal for providing feedback and support for this project.}  (IEE-GSEM University of Geneva, BREAD, CEPR) \\
   Enrico Moretti  (UC Berkeley, CEPR, NBER) \\
  Harrison Wheeler (University of Toronto) }

\begin{document}

\maketitle	
	

\vspace{2cm}
\begin{abstract} 
We  study the effect of localized housing price hikes on renters' mobility, consumption, and credit outcomes. 
Consistent with a spatial equilibrium model, we find that the consumption responses vary greatly for movers and stayers. While movers increase their consumption, purchase homes, and cars, stayers are relatively unaffected.
\end{abstract}


\thispagestyle{empty}

\newpage
\section{Introduction}

A growing empirical literature finds that changes in house prices  have sizable  effects on homeowner consumption and standard of living.  For example, \cite{mian2013household} and \cite{cerutti2017housing} find that the Marginal Propensity to Consume (MPC) out of housing wealth is 5-7 cents on the dollar, suggesting a significant wealth effect.\footnote{\cite{berger2018house} provide a model that  explains the magnitude of the effects.} 

Much less is known about the effect of  house price, and therefore rental cost, increases on renters. Rent increases in the years leading up to the Great Recession and more recent rent increases since the end of the Great Recession have generated widespread policy concerns in many cities about gentrification and renter displacement. For example, between January 2012 and 2018 the average rent in US cities increased by more than 20\%. In San Francisco, Seattle, and Boston, rents have increased on average by 40\%, 40\% and 22\%, respectively, with even larger increases in some gentrifying neighborhoods.\footnote{U.S. Bureau of Labor Statistics, Consumer Price Index for All Urban Consumers: Rent of Primary Residence in U.S. City Average, retrieved from FRED, Federal Reserve Bank of St. Louis, April 10, 2024.} Rents in San Francisco's Mission district and in Seattle's Pioneer Square area have increased by more than 50\% in the same period.\footnote{\cite{guerrieri2013endogenous} investigate which neighborhoods gentrify.  \cite{collinson2024eviction} study the effects of evictions on marginal tenants in Chicago and uncover limited effects on employment outcomes. Yet, there is limited systematic empirical evidence on the how rent increases affect incumbent renters consumption and standard of living. \cite{desmond2012eviction} provides a narrative  account of the effects of eviction on the lives of low income renters.}

In this paper, we  use a panel from Experian Credit Report Data on a 1\% sample ---or 2.2 million  individuals--- of the US population, with credit data longitudinally linked  between 2004 and 2016 to analyze renters' responses to house prices shocks.  We focus on non-housing consumption responses --- overall consumption, as measured by credit cards balances, home purchase, auto purchase ---and mobility responses.  In our analysis, a renter is exposed to a price shock if she lives in a zip code that experience a mean housing price increase of 25, 30, 50 percent over the three previous years. 

We find that, on average, the consumption effect of a price shock on renters  is not large. Renters who live in affected zip codes do not appear to change their overall consumption levels compared to renters of similar age who live in non-affected zip codes in the same CZ, nor do they seem to change the probability of having a mortgage or owning a car.  This is consistent with a spatial equilibrium model in which rent shocks are offset by wage increases, as would be in the case in which rents increase in some areas because of labor demand shifts, and local rents fully adjust  to higher local wages.  

However, we uncover vastly different effects when we split the sample between  ``movers''---defined as those who change address following the price shock---and stayers. While stayers experience modest declines in consumption, movers experience significant   increase in consumption, purchase homes and cars. Most of the movers appear to move to zip codes with lower housing prices.  A higher level of non housing consumption  is consistent with lower expenditures on housing. 

Overall, we conclude that the standard of living of renters in areas that are hit by house price increases do not decline significantly. If anything, they improve for those who decide to move.  While we have little to say about welfare and utility changes, our finding suggest that popular concerns about gentrification and renter displacement need to take into account endogenous wage changes for those who stay and lower costs of living for those who move. 
  
The reminder of the paper is organized as follows. Section \ref{sec:data} presents the data used.  Section \ref{sec:theory} introduces a simple conceptual framework. Section \ref{sec:descriptives} covers some descriptive results, while in Section \ref{sec:results}
we develop our empirical analysis. Finally, Section \ref{sec:conclusions} concludes.


\section{Data}
 \label{sec:data}

We use several data sources for our analysis, of which the main ones are: 1) credit report data for a 1\% panel ($ \approx 2.2m $ individuals) of the US population with credit score from 2004 to 2016; and, 2) house price data series from the FHFA and other sources. Here, we describe the two sources of data and their limitations.

\subsection{Credit Data}

The source of the credit report data is Experian.  
Selection into our sample depends on having a valid credit score at a point in time (in 2010).  Over 70\% of the US population has a credit report in 2010, with that share increasing over age.

A credit report is generated any time an individual opens or initiates a credit operation, e.g. applies for a credit card, a car loan, a mortgage, etc. In general, any formal credit operation will generate a record in the report. We draw our data on June 30th of every year, from 2004 to 2016, and at that date we register all the closing balances transmitted by the financial institution to Experian.\footnote{From Experian source: credit card lenders don’t typically report at the end of the month. They report throughout the month in cycles, every single day a portion of the portfolio. Experian takes the snapshot at the end of the month so to take into account everything that has been reported. For example,  most credit card lenders report the statement balance that was sent out on the consumers' monthly invoice or statement. Typically, for credit cards that means that even for those customers that pay in full every  month, the balance reported would be the full cycle expenditure.}

An  advantage relative to data like Mint.com is the fact that selection into the sample does not depend on user sign-up.
An important limitation of our sample is that we miss those individuals who have no valid score in 2010, these are individuals with no credit lines (sometimes referred to as unbanked) or insufficient credit information -- overall, accounting for no more than 20\% of the adult population.\footnote{\url{https://files.consumerfinance.gov/f/201505_cfpb_data-point-credit-invisibles.pdf}}
Both groups are likely to be over-represented among low-income households. 
A second limitation is that we do not observe family or household structures, as our data are drawn at the individual-level. 

For each individual--year pair in the sample,  we have information on residential location at the level of zip code, year of birth, (imputed) income,  credit use, and delinquencies. Specifically, we have information on all individual credit lines and balances: credit cards,  mortgages, auto loans and personal loans. Credit card balances are obtained by Experian from the different credit card companies.  In addition, we have information on delinquencies, defaults, as measured by  balances outstanding after 1, 2, 3, 4, 6 months, and  chapter 7 and 13 status and foreclosures.

To have a sense of how our sample compares with the US population,  Appendix Figure \ref{fig:a1} plots the log size of our sample in each commuting zone against the log number of households in the 2010 ACS. Two features are noteworthy. First, there appears to be a tight link between our sample size and the corresponding number of households in ACS. Second, the slope (standard error) is essentially 1 (0.006 or 0.007 if we weight by population), the Experian data somewhat  under-sample lower population/rural areas and over-sample larger cities.  

Appendix Figure \ref{fig:a2} plots the log size of our sample in each zip code against the log number of households in 2010 IRS data. The slope (standard deviation) here is again very close to 1 (.013). The Figure also reveals significantly more dispersion in zip codes that are small according to IRS data.   
In essence, our sample tends to underrepresent zipcodes with fewer than 400 IRS returns, but appears to be quite tight for any zipcode with more than 1800 returns.

Throughout the paper, we define as homeowners those who have a mortgage, and as renters those without a mortgage. This definition is likely to contain measurement error, since we miss ownership cases when a property is paid in cash or the mortgage is fully repaid.    However, we expect that the probability of having fully repaid a mortgage is low in the age group that we are focusing on: 25-55 years old. Similarly, cash purchases might be less relevant in the prime age segment.\footnote{The share of houses sold in cash in the period of interest appear to be within .03 and .12, U.S. Census Bureau and U.S. Department of Housing and Urban Development, Houses Sold by Type of Financing, Cash Purchase [HSTFC], retrieved from FRED, Federal Reserve Bank of St. Louis; \url{https://fred.stlouisfed.org/series/HSTFC}, April 19, 2024. We have no information on the break-down by  age of the buyer.}

Appendix Figure \ref{fig:a6} plots the probability of homeownership over the lifecycle in our data and in the 2010 American Community Survey (ACS). The Figure indicates that we are indeed underestimating ownership rates, especially at younger ages. This is perhaps because those owners purchase their homes through parental transfers. What is reassuring though is that by age 33 the two lines are quite close, the differences are of 2/3pp, and the shapes are very comparable.

Furthermore, Appendix Figure \ref{fig:a7} plots ownership rates in our data against ownership rates in the 2010 ACS. We then estimate  the slope (standard deviation) of that relationship to be .46 (.065). Measurement error in ownership status does not appear to have any obvious geographical component.


{\bf Income} We have access to two  measures of income: 1) An individual W2 based measure, and 2) a more comprehensive measure \textit{Total Income} which is based on W2 and other sources of income (e.g. investment, farming, and so on).\footnote{It is important to note that the reason why those measure are available is because of ``market'' need for detecting the ability to pay by potential borrowers to be incorporated in the credit score and credit applications. Below is an extract from the Experian document on the matter:
...Final Federal rules implementing the Credit CARD Act became effective on February 22, 2010, require that card issuers consider a borrower’s ability to pay in lending decisions.  This allows use of empirically derived and statistically sound models Income Insight released December 2009
New rules issued on March 18, 2011
Require card issuers to consider borrower's independent income, rather than household income New rules become effective on October 1, 2011. Income Insight W2 released April 2011. \textbf{Income Insight} provides a comprehensive measurement of total income offering a complete financial picture of a borrower for greater insight into their ability to meet their obligations.
\textbf{Income Insight W2} estimates the wage income of a consumer which is especially applicable for low to mid-range income levels where income is comprised mostly from wages.
Both measures are top coded at 900,000USD, and are ultimately significantly highly correlated, $.8332$.}

Our measure of incomes are inputed and validated through third party data. While the exact methodology of inputation is proprietary to Experian, we do know the broad steps taken for producing such measures. In practice, all the variables available in the credit report data and premier attributes (400 variables characterizing credit behavior) are used, while demographic information is not used. Experian had access to a number of administrative records on income  either as W2 or as total incomes for the years 2005 to 2009, then validated subsequent waves for a smaller sample. From that, they constructed an imputation model. The model appears to perform relatively well and is able to match fairly closely some moments of the income distribution circulated by the IRS.

Like any imputation, these are likely to be measures of income with error.  The Appendix Figure \ref{fig:a3} shows average income, at the zip code level, versus IRS derived ones. Appendix Figure \ref{fig:a4} plots the life cycle profiles of incomes from Experian and the ACS. While, as expected given selection into the credit market, the Experian imputed income appears higher than the ACS one, the shape of the profiles are very comparable. To get a better sense of the representativeness of our sample, Appendix Figure \ref{fig:a3} plots mean income in our sample in each zip code against household income in the IRS data. The slope here is .4 (.004), indicating that the Experian income tracks IRS income.

In our analysis, we focus on individuals aged 25 to 55 years old. Appendix Figure \ref{fig:a4} indicates that mean income by age in our data track reasonably well to mean income by age in the  ACS, in the prime age range. While income in the data is higher in each age group, income changes over the life cycle appear to be consistent with ACS data.

Overall, despite the limitations of our data, the income distribution in our data appears to match well with the income distribution from the ACS micro-level data and its changes over the life cycle.

{\bf Consumption.} We create measures of consumption  using information on  auto loans balances, mortgage balances, and credit card balances. 
For auto loans and home mortgages, we observe both loan origination and amount of the loan. Thus, we can measure both the probability of buying a car and a home (the extensive margin) as well as the amount spent on cars and homes (the intensive margin).\footnote{While it is true that car loans and mortgages balances and payment will reflect the terms of the loans, those are indeed the amounts paid by the consumers at the end of the loan cycle, including the interest rate payments.}  These measures of consumption are not perfect because they are based on loans. Thus, we miss car and home purchases made in cash, and of course we miss the component due to down payments.   

To measure consumption of goods outside cars and real estate, we use information on credit card balances. \cite{collinson2024eviction}  and \cite{diamond2021standard}  take a similar approach.
The idea is that for consumers who regularly pay their credit card balance, their  credit card balance, on June 30th, is a good measure of the value of goods purchased by credit card in a representative month.  
The problem of course is that for consumers who do not  pay their full credit card balance regularly, the balances include not just the value of goods purchased, but also  interest payments (and other  fees). In our data, interest charges and fees are not separately identified. One, could however argue that such fees are in fact the consumption of financial services. The snapshot of data on card balances will in general give the end of cycle balance of credit card, i.e. the amount spent in the latest cycle.

In our analysis of credit card purchases, we focus on credit card balances conditional on the consumer not experiencing a large negative decline in her credit score in the sample period and not defaulting on her credit card debt by 2016, which is the end of the sample period. Negative declines in her credit score are  drops of 100 points in any year of the data.  We also provide a separate analysis where we focus on  the probability of experiencing large negative decline in credit score and the probability of default as outcomes.  

The assumption is that consumers who do not experience large negative decline in credit score and by the end of the sample period have never defaulted, tend to pay their monthly balance regularly,  or if they accumulate debt at some point during the sample period, they tend to repay it in a timely fashion. 
This assumption fails for consumers who accumulate debt  in the absence of a significant downward movement in their credit score and in the absence of default. 
This type of measurement error seems unlikely to be widespread. A large unpaid balance is likely to trigger a downward movement in credit score and, in some cases, it may result in default. However, it is possible that this measurement error is present in the last year of our data, since for some consumers whose balances include significant interest charges in late 2016, the credit score may not have adjusted  because those charges are too recent to trigger a credit score adjustment within 2016.  


Under our assumption,  credit card balances can be used to measure consumption by credit card. 
Note however that even for consumers for whom this assumption is valid, there might be measurement error in  the exact timing of consumption. For example, if a consumer accumulates debt in 2010, and repays it in 2012, we would overestimate consumption in 2011 and underestimate in 2012.

Since this paper is about consumption, it is important to benchmark our measure of consumption based on credit card balances against other known measures of consumption. As a first benchmark, we compare our measure of average household expenditure with the corresponding measure in the National Income and Product Accounts (NIPA) and in the Consumer Expenditure Survey (CEX). To make NIPA data comparable to our data and to CEX data, we remove from NIPA receipts from sales of goods and services by nonprofit institutions and net expenditure from retirement plans. This figure shows that our measure of consumption expenditure appears to match NIPA aggregate consumption  closely, although not perfectly. By contrast, average consumption in the CEX is only 71\% and 75\% relative to NIPA and our data, respectively. This is consistent with the fact that CEX has high non-response among high income households \cite{sabelhaus2000can}. Even more importantly, the CEX significantly underestimates consumption relative to NIPA \cite{attanasio2014consumption}, \cite{aguiar2015has}. We also note that the average savings rate in our data is not very different from that in NIPA at 5.5\%.  Overall, our data appear to be more comprehensive than the CEX, at least in comparison with aggregate expenditures in NIPA.

A separate issue is the fact that  consumption expenditure can take many different forms. In particular, consumers may pay using credit card, debit card, ACH, cash and checks. This matters for us because our data provide different amounts of information on the type of consumption based on the form of payment. While we observe the sum of all expenditure by credit card, we do not observe expenditures by other means. 
On average, 31\% of expenditures in 2022 are by credit card. This share varies across income groups, but not very much (2023 Federal Reserve Report).\footnote{\url{https://tinyurl.com/yc3n9bvh}} For example, it is 20\% for consumers with income between 25,000 and 50,000,  31\%
for consumers with income between 75,000 and 100,000, and 37\% for consumers with income between 100,000 and 150,000.  In our empirical analysis, we will assume that this share is uncorrelated with exposure to price shocks. 
More specifically,  we assume that the share of expenditures paid by credit card by a specific individual is not systematically associated with the rent paid by that individual.

\subsection{House Price Data} 
\label{sec: shocks}

We measure house price shocks using  data at the zip code level for years 1998 to 2016. 

\begin{figure}[ht!]
    \centering
    \begin{minipage}{0.3\textwidth} 
        \centering
        \includegraphics[scale=.2]{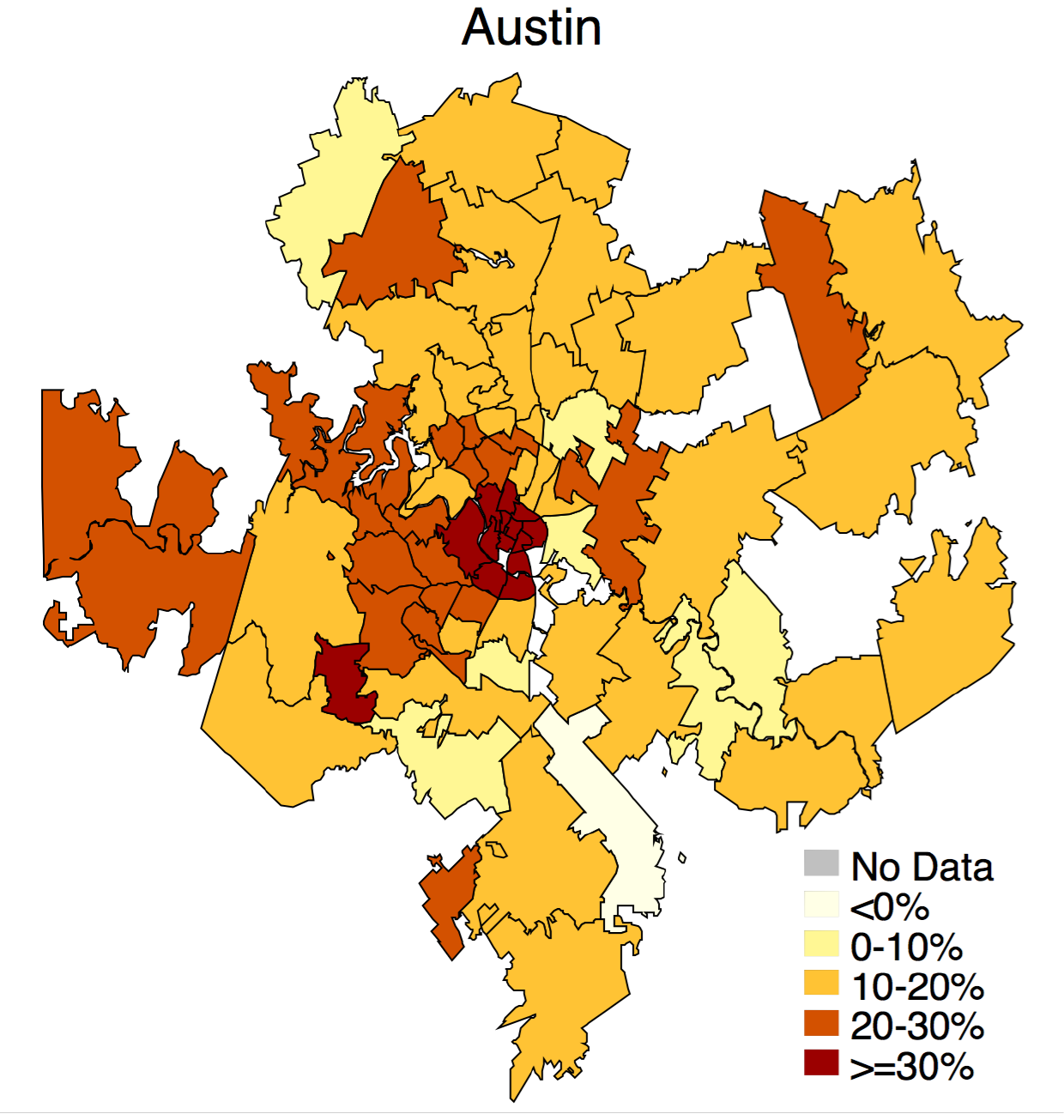} 
        
        \label{fig:austingr}
    \end{minipage}
    \hfill
    \begin{minipage}{0.3\textwidth} 
        \centering
        \includegraphics[scale=.2,angle=-90]{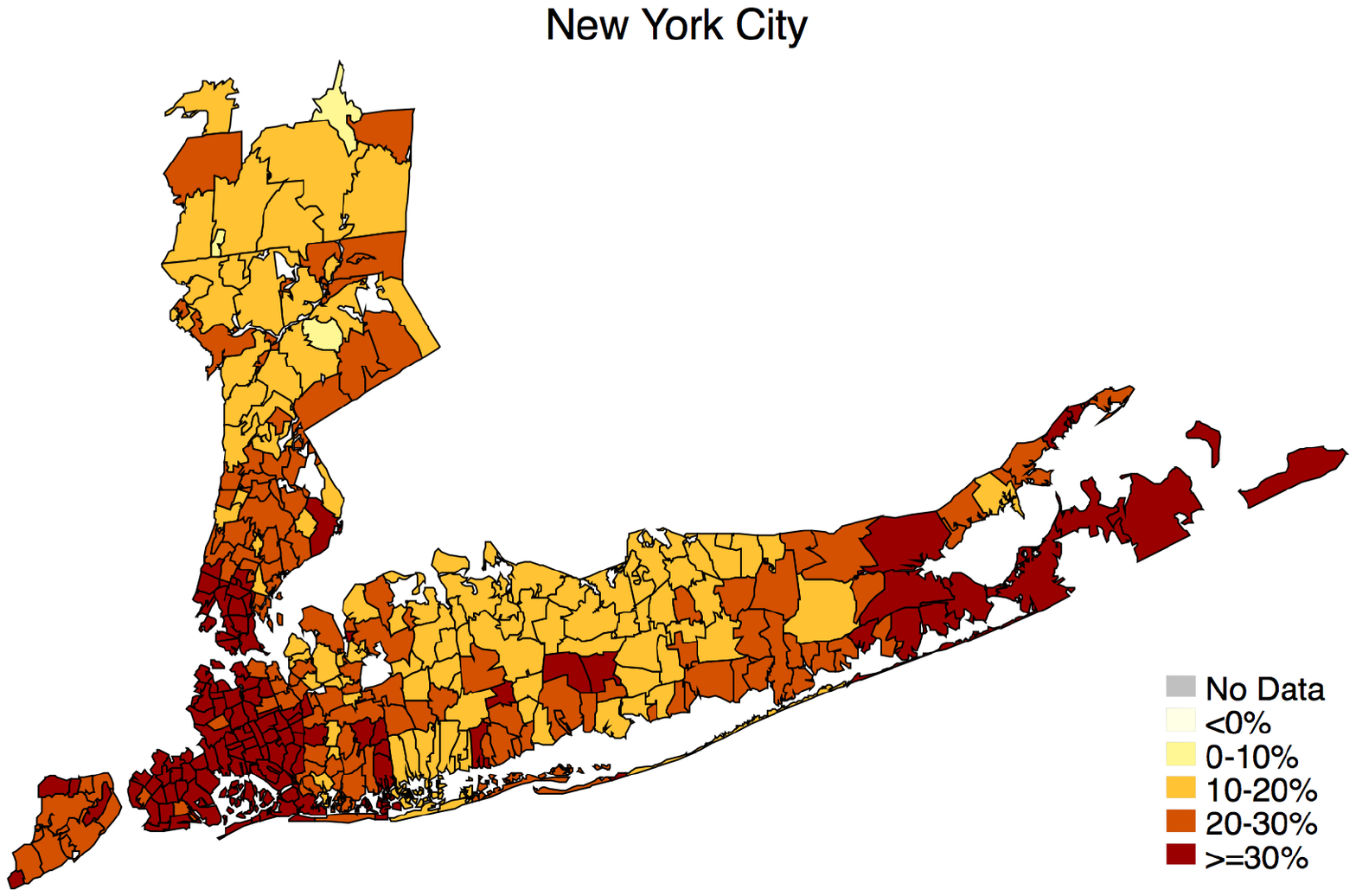} 
        
        \label{fig:nycgr}
    \end{minipage}
    \hfill
    \begin{minipage}{0.3\textwidth} 
        \centering
        \includegraphics[scale=.2]{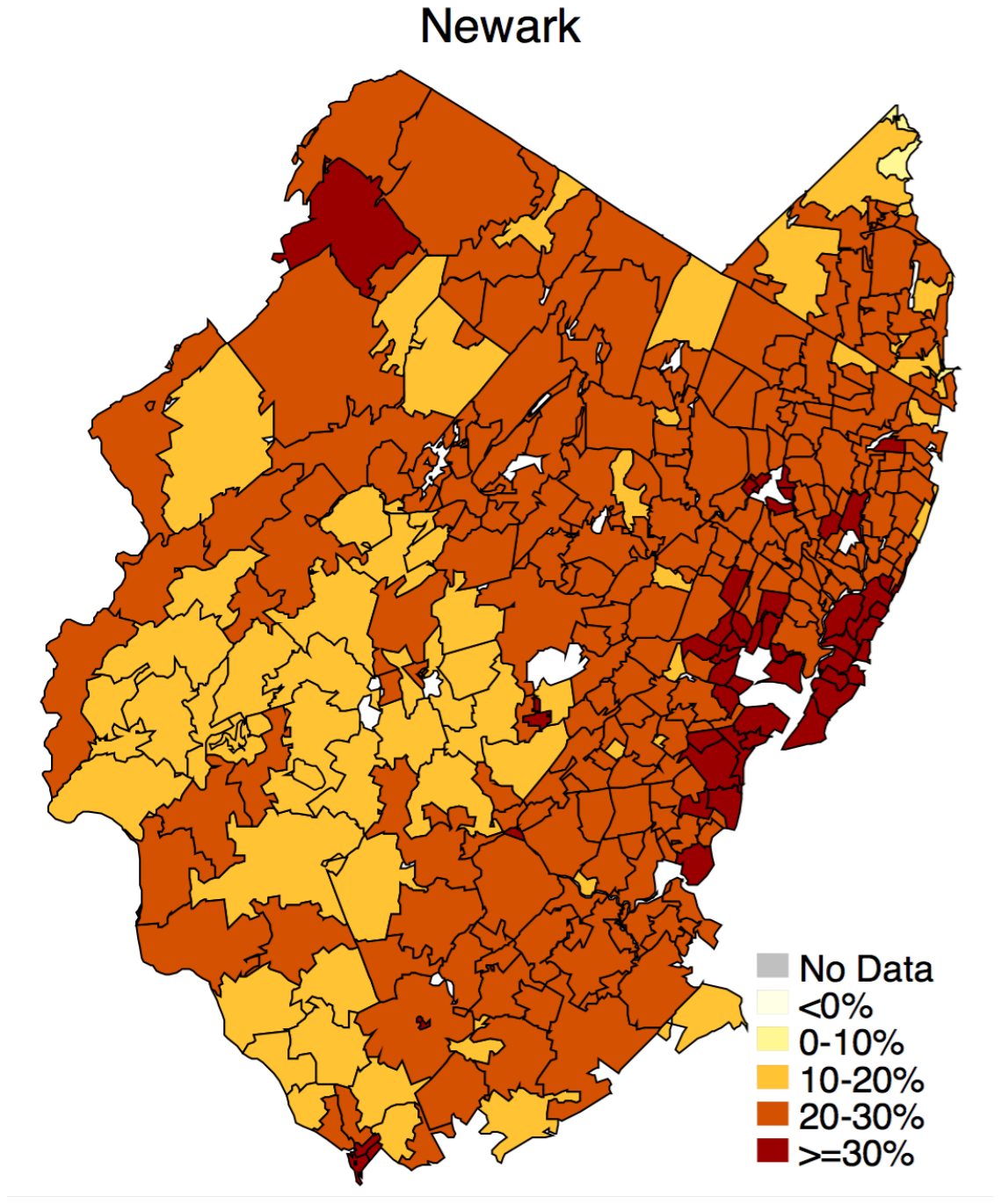} 
        
        \label{fig:newarkgr}
    \end{minipage}
    \\
    \begin{minipage}{0.3\textwidth} 
        \centering
        \includegraphics[scale=.2]{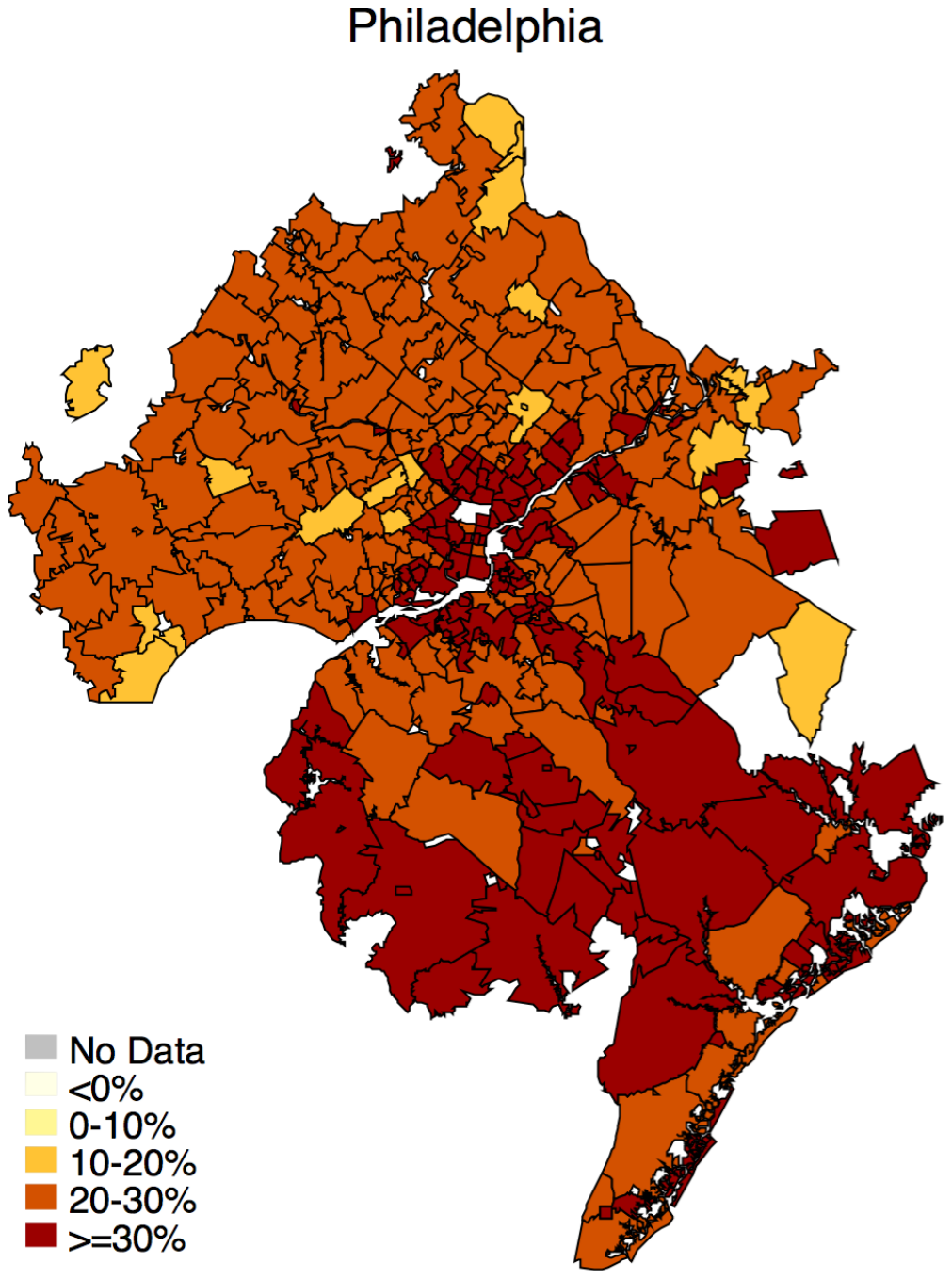} 
        
        \label{fig:phillygr}
    \end{minipage}
    \hfill
    \begin{minipage}{0.3\textwidth} 
        \centering
        \includegraphics[scale=.2]{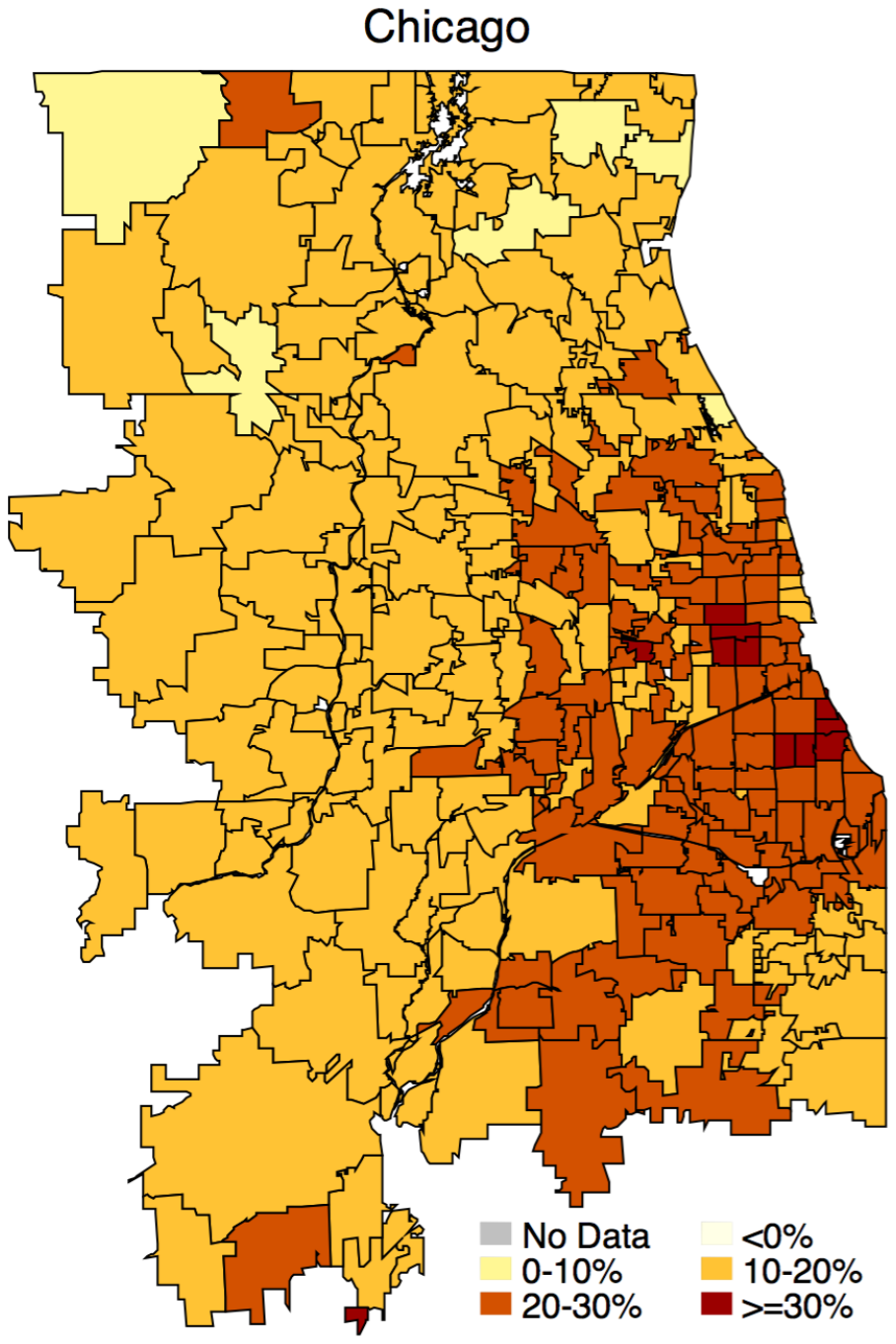} 
       
        \label{fig:chicagogr}
    \end{minipage}
    \hfill
    \begin{minipage}{0.3\textwidth} 
        \centering
        \includegraphics[scale=.2,angle=-90]{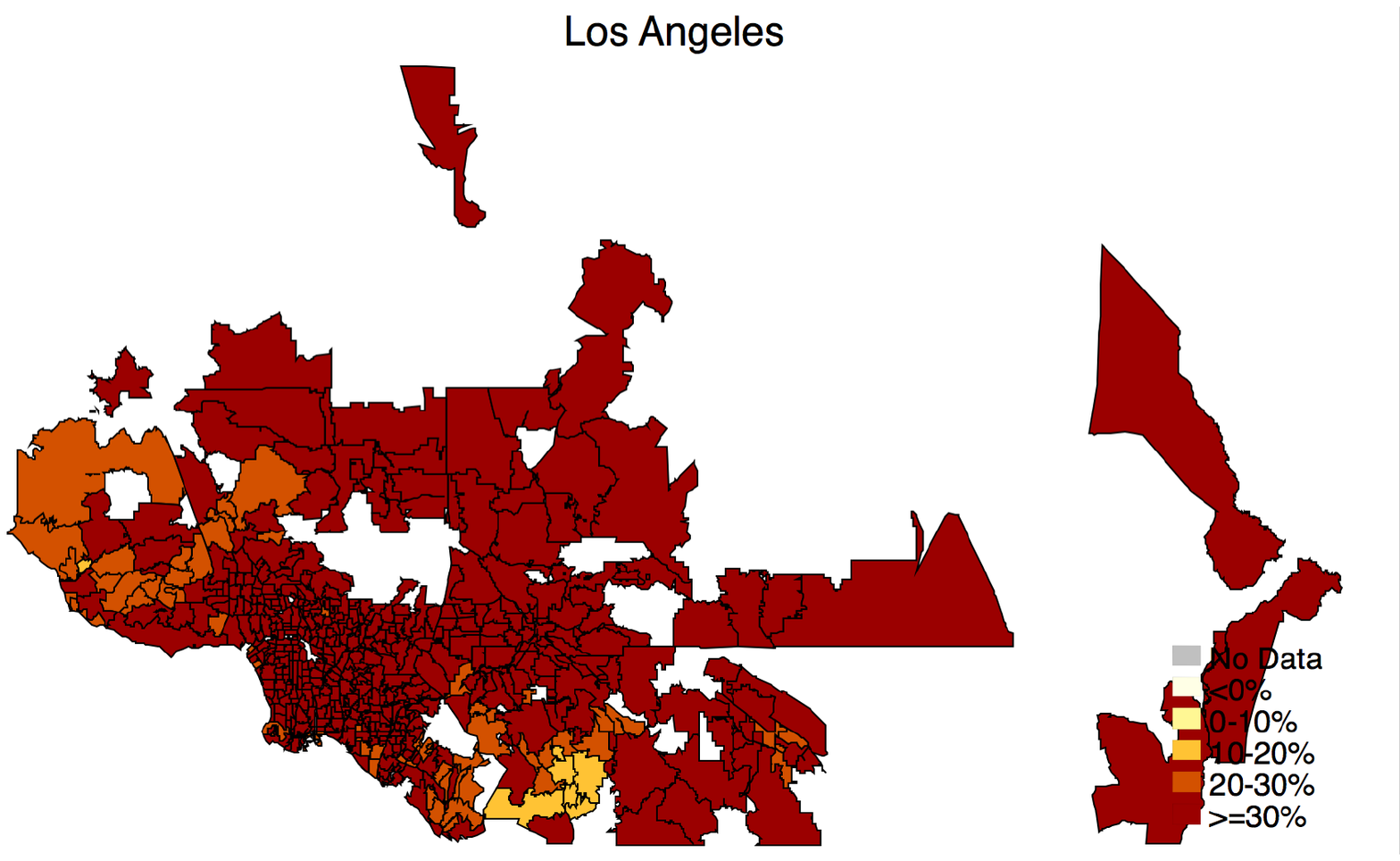} 
       
        \label{fig:lagr}
    \end{minipage}
    \\
    \begin{minipage}{0.3\textwidth} 
        \centering
        \includegraphics[scale=.2]{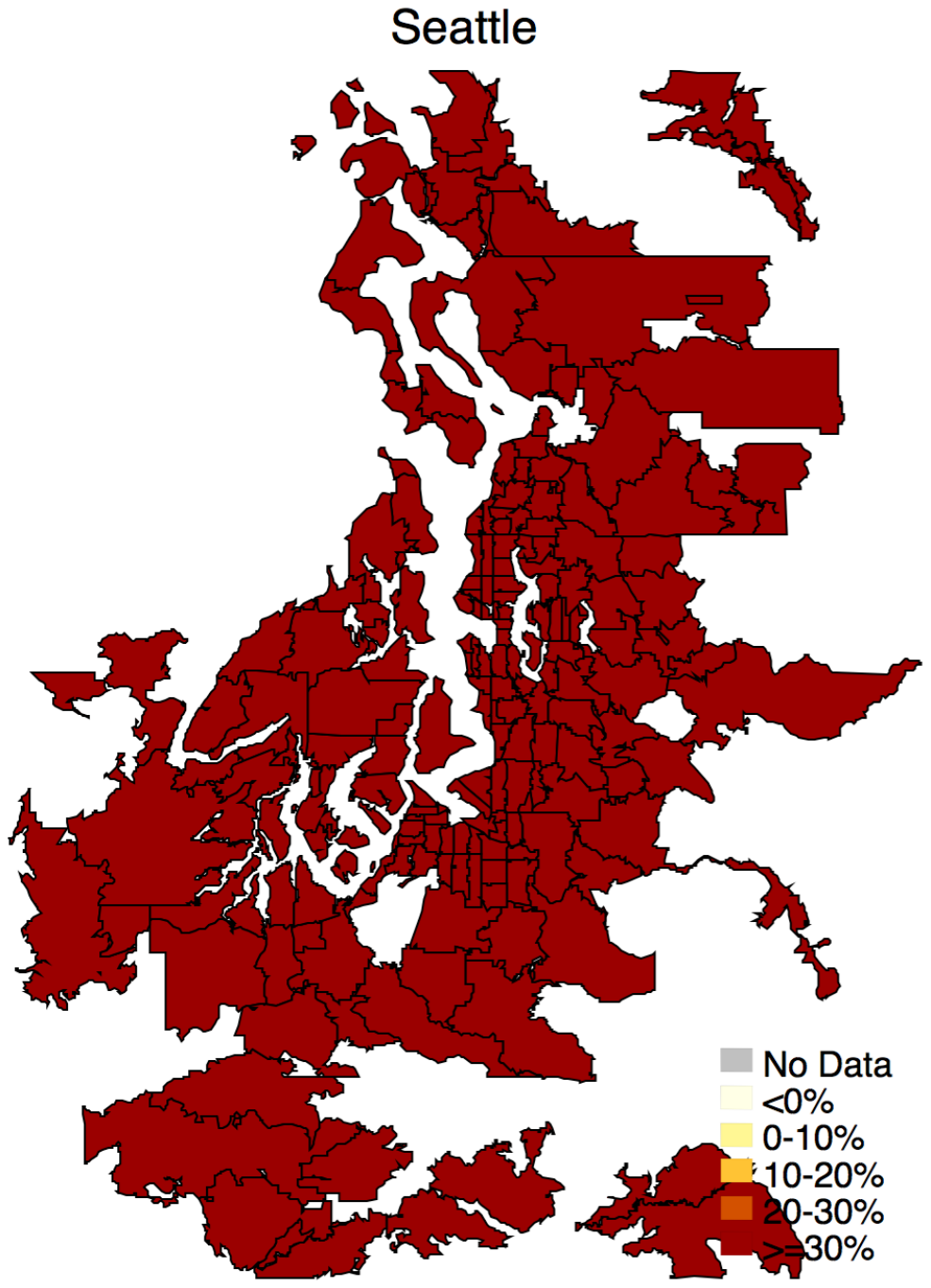} 
       
        \label{fig:seattlegr}
    \end{minipage}
    \caption{Examples of within cities shocks at zip code level in 2007}
    \label{fig:cities_percent}
\end{figure}

Appendix Figure \ref{fig:a8} and \ref{fig:a9} shows the national time series of price level.  More important for us,  Figure \ref{fig:cities_percent} shows some examples of   mean price changes at the zip code level in 2007 for some selected cities. 
 
\begin{figure}[ht!]
    \centering
    \begin{minipage}{0.3\textwidth}
        \centering
        \includegraphics[scale=.25]{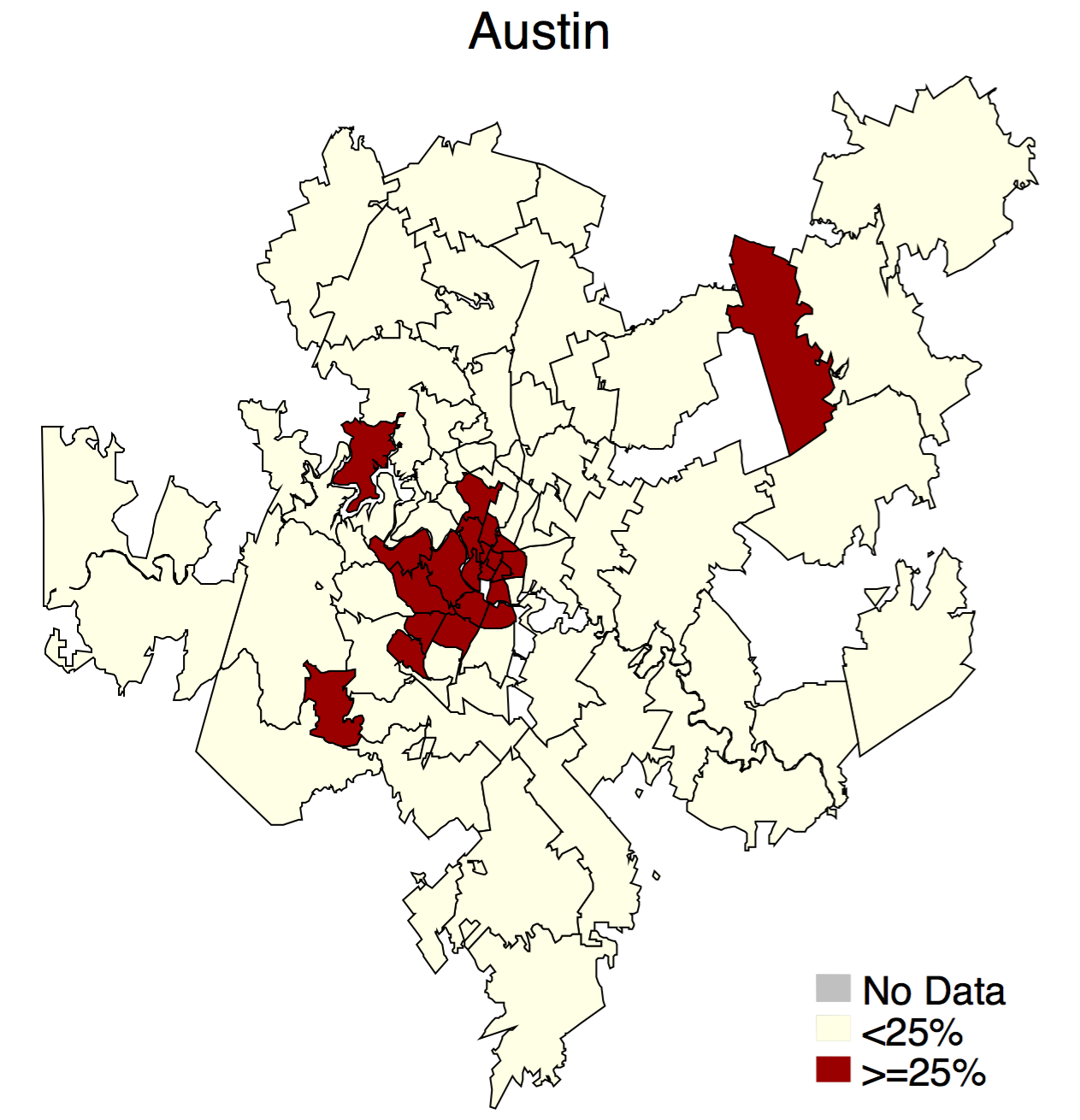}
        
        \label{fig:austin}
    \end{minipage}
    \hfill
    \begin{minipage}{0.3\textwidth}
        \centering
        \includegraphics[scale=.25]{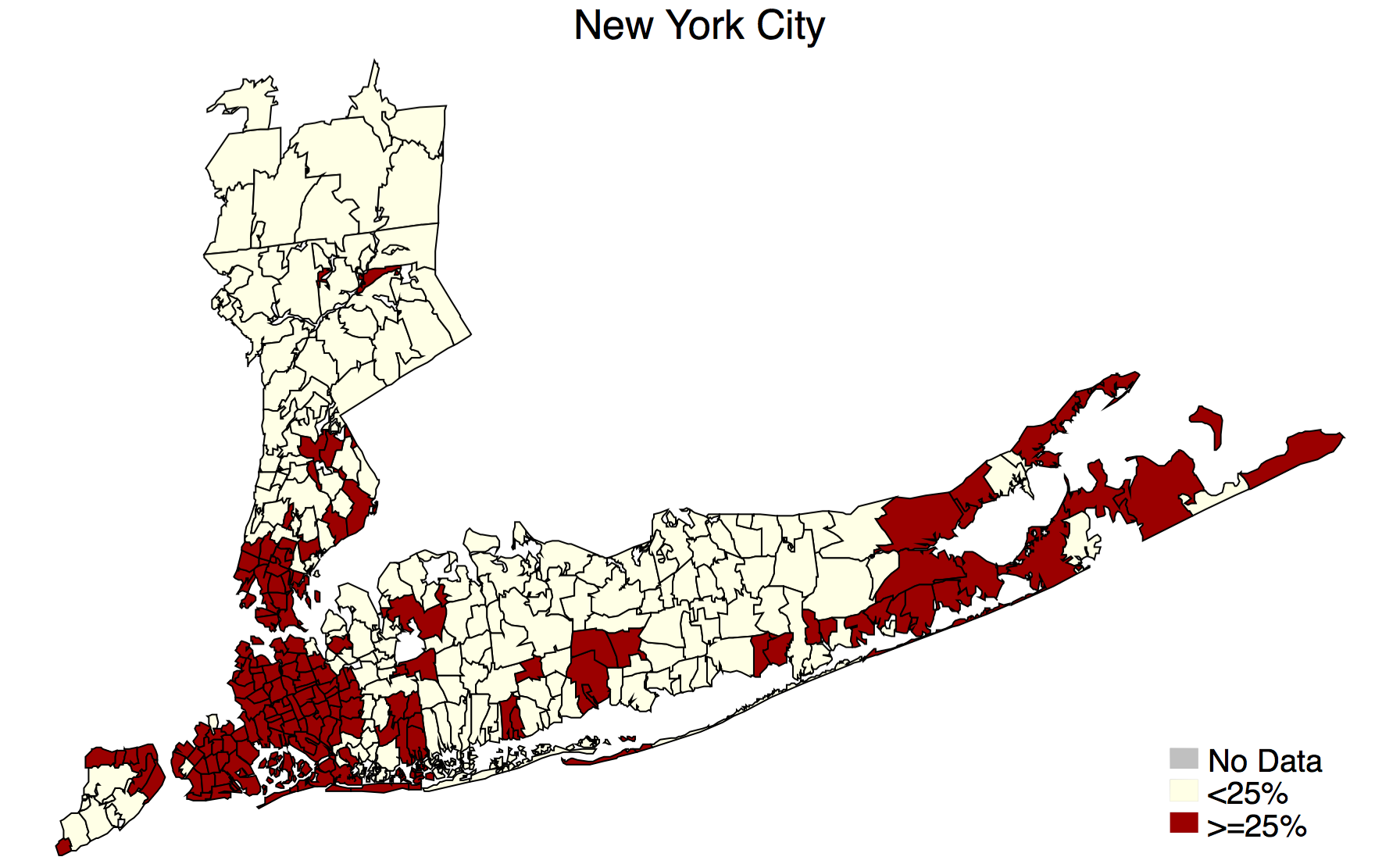}
        
        \label{fig:nyc}
    \end{minipage}
    \hfill
    \begin{minipage}{0.3\textwidth}
        \centering
        \includegraphics[scale=.25]{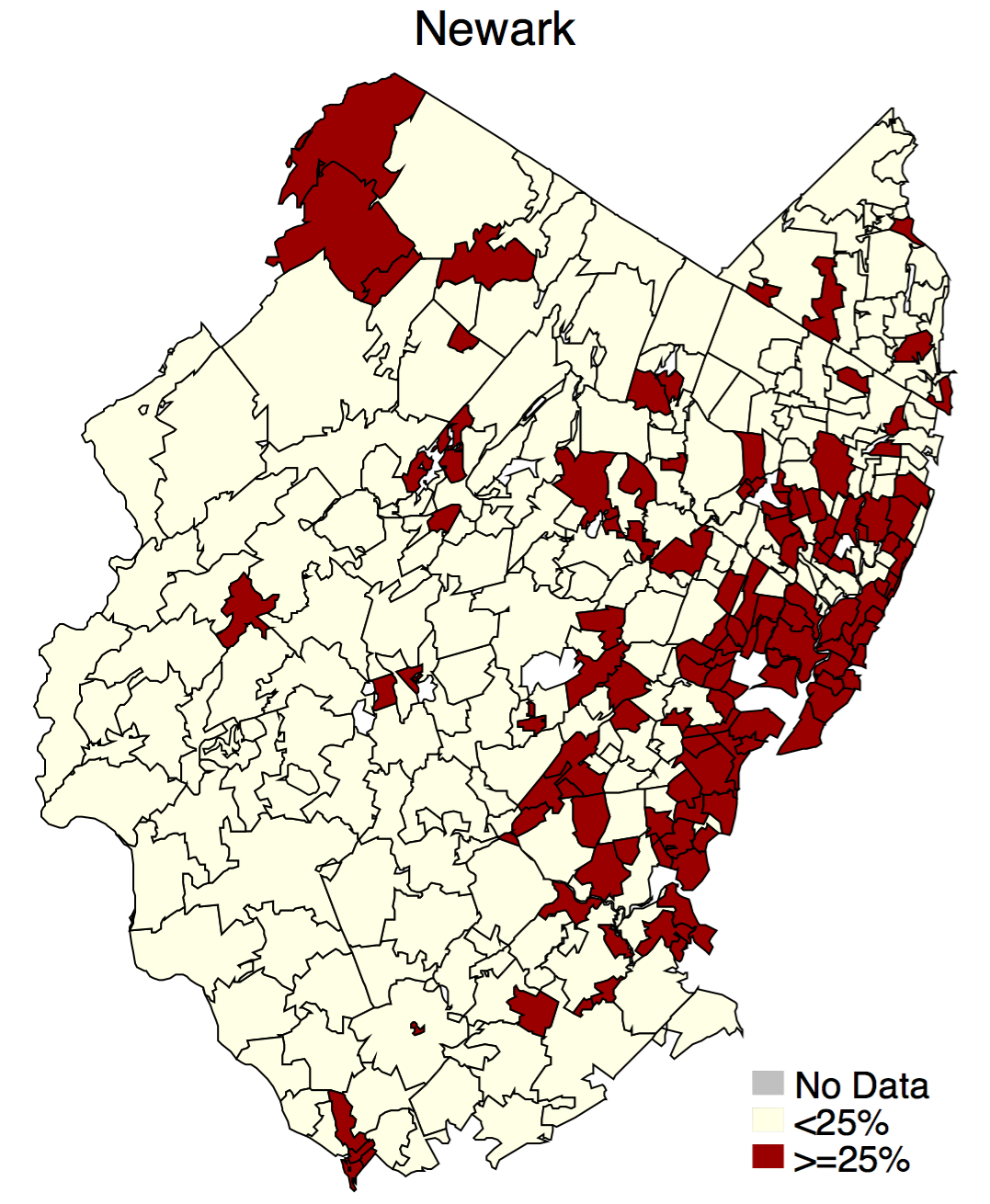}
        
        \label{fig:newark}
    \end{minipage}
    \\
    \begin{minipage}{0.3\textwidth}
        \centering
        \includegraphics[scale=.25]{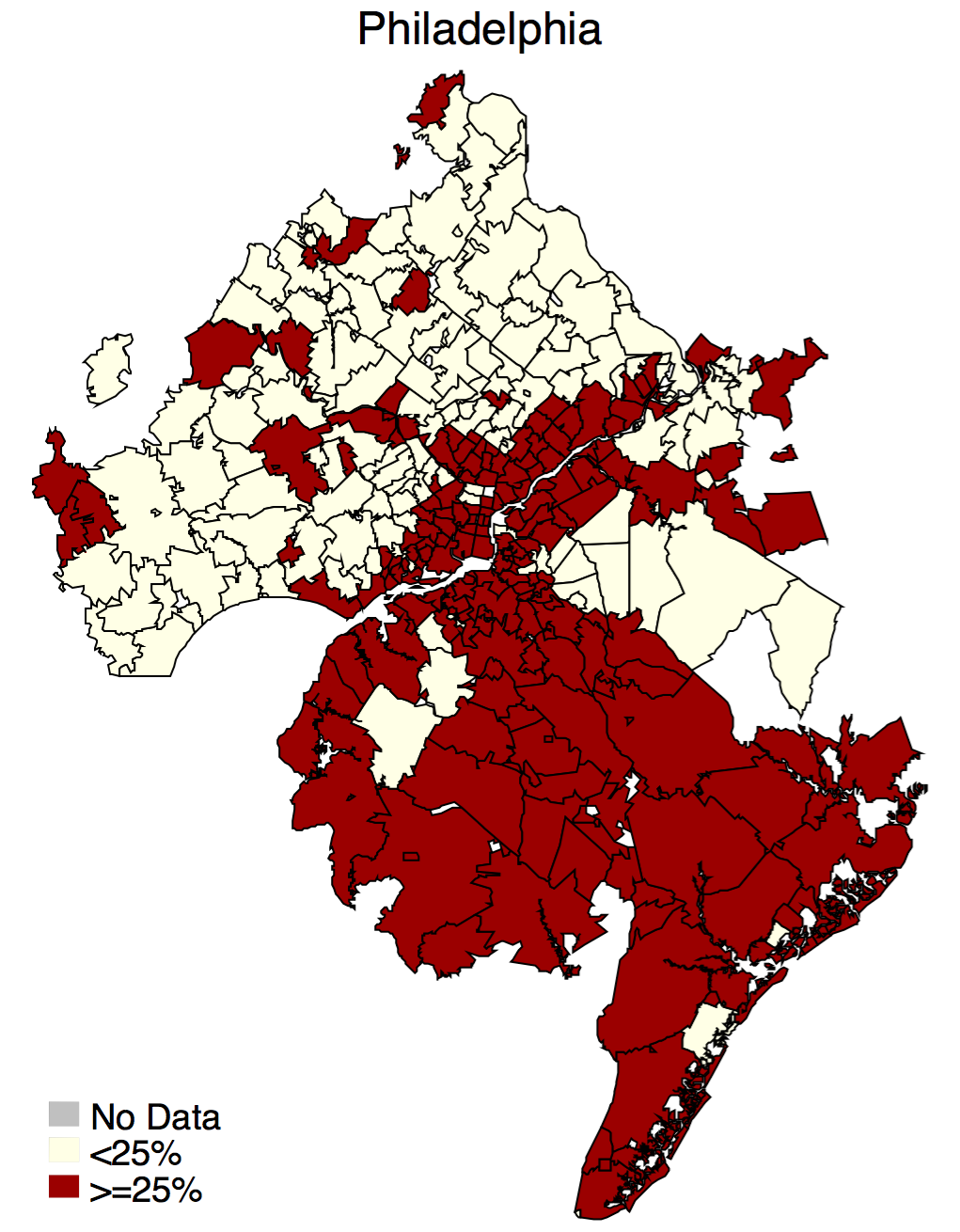}
        
        \label{fig:philly}
    \end{minipage}
    \hfill
    \begin{minipage}{0.3\textwidth}
        \centering
        \includegraphics[scale=.25]{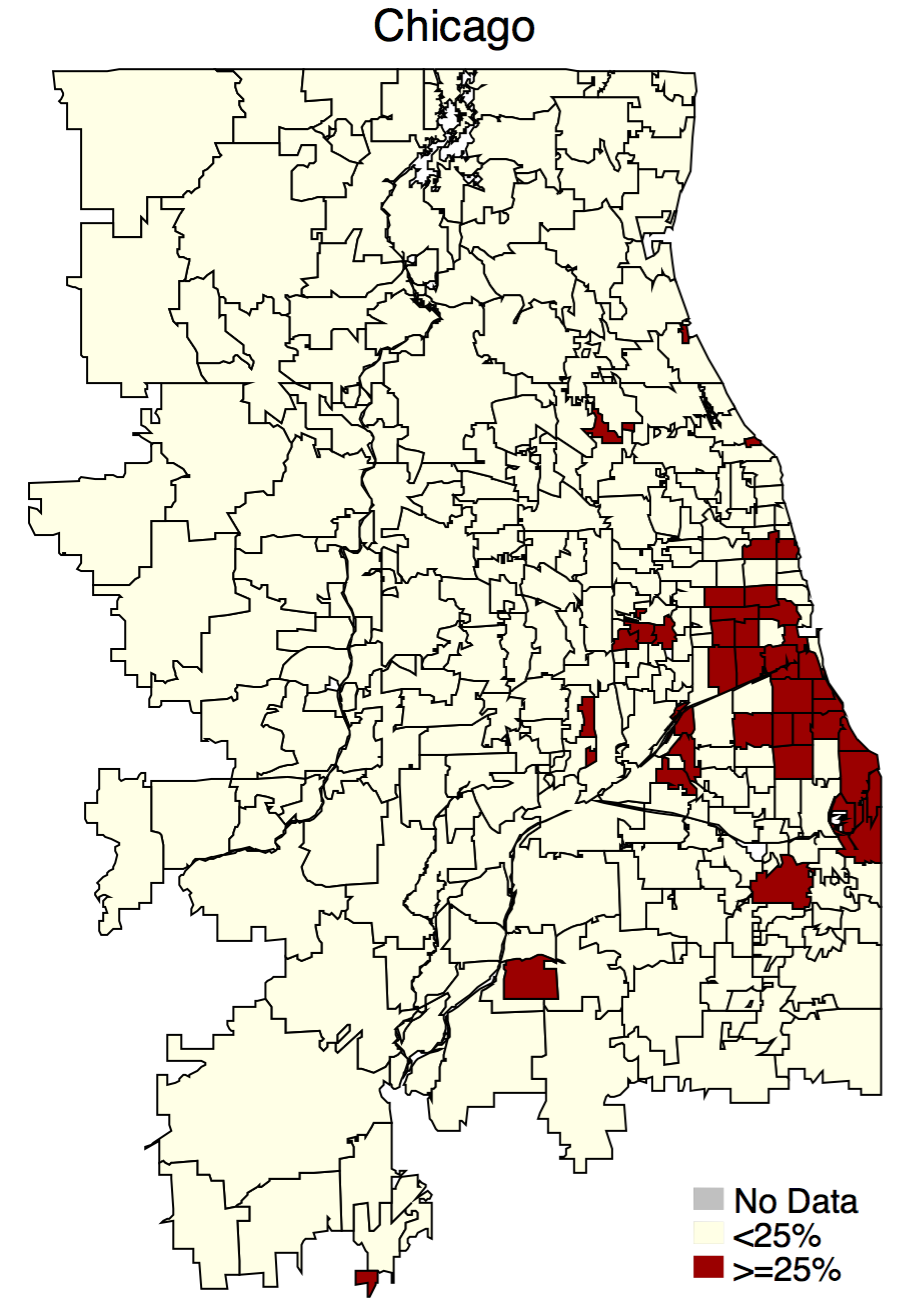}
       
        \label{fig:chicago}
    \end{minipage}
    \hfill
    \begin{minipage}{0.3\textwidth}
        \centering
        \includegraphics[scale=.25]{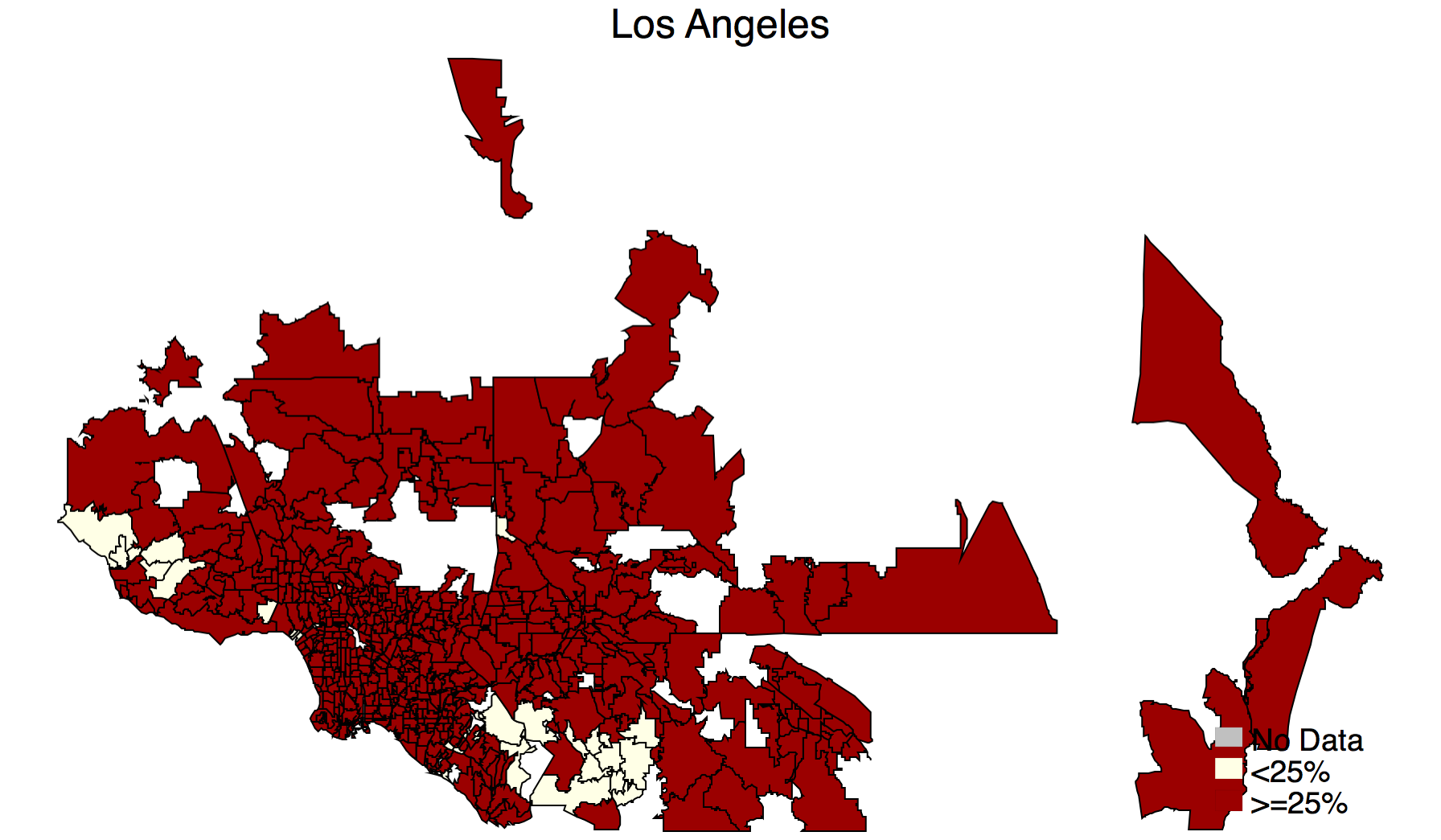}
       
        \label{fig:la}
    \end{minipage}
    \\
    \begin{minipage}{0.3\textwidth}
        \centering
        \includegraphics[scale=.25]{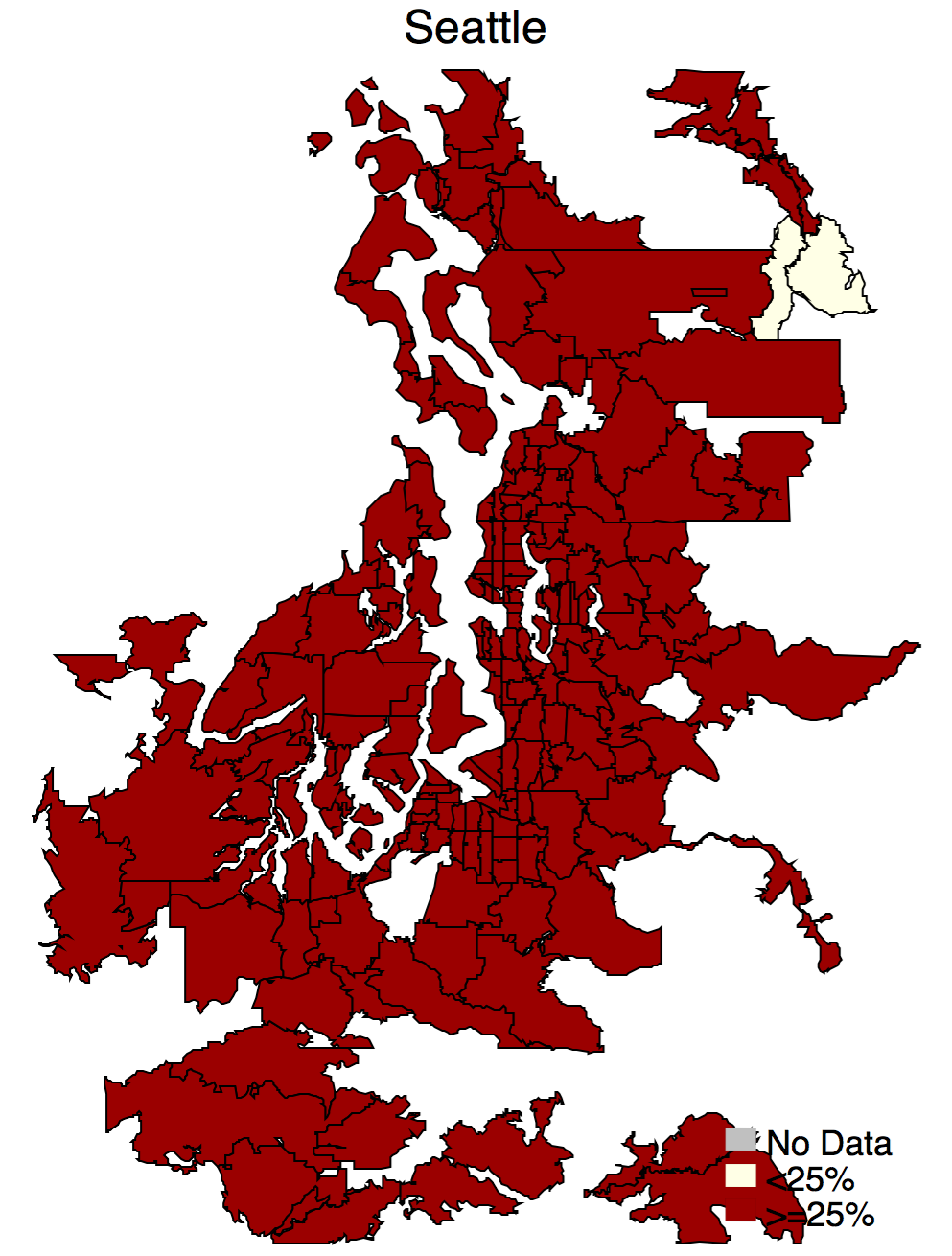}
       
        \label{fig:seattle}
    \end{minipage}
    \caption{Examples of within cities shocks at zip code level in 2007}
    \label{fig:cities}
\end{figure}

There is a large variation within Commuting Zones (CZs).  The geographical unit of analysis is the zip code.  We define as shocks to house prices those price increases of 25  percent over the three previous years in 2007 and 2014 at the zip code level (we also used 30 and 50 percent as thresholds). For 2007, Figure \ref{fig:cities}    shows which zip codes are affected in seven selected CZ: Austin, New York, Newark, Philadelphia, Chicago, Los Angeles and Seattle. The within CZ distribution of shocks is not the same in all cities. Rather, it appears to be specific to each city. For example, while shocks were concentrated near the city center in some cities, they were more spread out in others. 

Appendix Table \ref{tab:balance1} shows means of the key variables in the areas that experience a price shock and areas that do not. 
A comparison of ``treatment'' and ``control'' zip codes suggests that home prices increases occurred in areas with a younger population, lower credit score, and lower incomes. These differences are not economically  large, and are consistent with \cite{mian2013household} and \cite{albanesi2022credit}.  

Appendix Table \ref{tab:balance2} compares average characteristics of movers and non-movers. Not surprisingly, movers are
younger, have lower scores have slightly  lower income and consumption and higher car loan balances.
 
Since we focus on renters, a more direct measure of rent shocks would be changes in rents. We use house prices because unlike rents they are available at the zip code level for the entire country. 
In principle, changes in rents and house prices don't need to be perfectly correlated. Variation in rents reflects current housing demand and supply, while variation in house prices also reflect expectations of future appreciation   and depreciation, and credit market conditions.  To assess the relationship between rent changes and house price changes in our sample period, we use ACS data at the MSA level. 

Figure \ref{fig:a_rent} plots log house price changes  on the x axis and log rent changes on the y axis for the 300 MSAs from 2000 to 2010.  The figure shows that cities with an increase in house prices also experience increases in rents. The slope is 0.43 (.03), indicating that empirically a 10\% increase in house prices is correlated with 4.3\% increase in rents. This magnitude is useful in scaling our empirical estimates below.


\section{Framework}
\label{sec:theory}


This section presents a simple spatial equilibrium model of the labor market and housing market, which is useful for considering how shocks to cost of housing may impact different individuals.  
We adopt the standard assumptions of Rosen-Roback spatial equilibrium models, with specific functional form assumptions similar to those in \cite{moretti2011}.  For brevity, we focus on the simplest version of the model with intuitive closed-form solutions (\cite{moretti2011}, \cite{klinemoretti}).

There are two cities, $a$ and $b$.  Each city is a competitive economy, producing a single output good $Y$ that is traded on the international market at a fixed price normalized to 1.  The production function in city $c$ is: $\ln Y_{c} = A_{c} + (1-h) n_{c}$, where $A_{c}$ is city-specific log total factor productivity (TFP);  $n_{c}$ is the log of the share of employment in city $c$; and $0 < h < 1$.  

Indirect utility of worker $i$ in city $c$ is given by: 

$v_{ic} = w_{c} - \beta r_{c} + x_{c} + e_{ic}$ 

where $w_c$ is the log of nominal wage, $r_{c}$ is the log of cost of housing, $x_{c}$ is the log value of amenities, and $\beta$ measures the importance of housing consumption in utility and equals the budget share spent on housing.  Since people do not spend their entire budget on housing, the effect of a 1\% increase in rent is smaller than the effect from a 1\% decrease in wage.

The random variable $e_{ic}$ is an idiosyncratic location preference, for which a large draw of $e_{ic}$ means that worker $i$ particularly likes city $c$ aside from real wages and amenities.  We assume that worker $i$'s relative preference for city $b$ over city $a$ ($e_{ib} - e_{ia}$) is distributed uniformly $U[-s,s]$.  The assumption of a uniform distribution is analytically convenient, allowing us to derive closed-form expressions for the endogenous variables in equilibrium.  The comparative statics are unchanged in an extended version of this model that assumes the $e_{ic}$'s are distributed according to a type I Extreme Value distribution \cite{klinemoretti}.

Workers locate wherever utility is maximized.  Worker $i$ chooses city $b$, rather than city $a$, if and only if the strength of location preferences exceeds any real wage premium and higher amenity value: $e_{ib} - e_{ia} > (w_{a} - \beta r_{a})  - (w_{b} - \beta r_{b}) + (x_{a} - x_{b})$.  In equilibrium, there is a marginal worker who is indifferent between city $a$ and $b$.

The parameter $s$ governs the strength of idiosyncratic preferences for location and, therefore, the degree of labor mobility and the city's elasticity of local labor supply.  If $s$ is large, many workers will require large differences in real wages or amenities to be compelled to move, and the local labor supply curve is less elastic.  If $s$ is small, most workers are not particularly attached to one city and will be willing to move in response to small differences in real wages or amenities, and cities face a more elastic local labor supply curve.  In the extreme case where $s$ is zero, there are no idiosyncratic preferences for location and there is perfect labor mobility.  In this case, workers will arbitrage any differences in real wages adjusted for amenities and local labor supply is infinitely elastic.

We characterize the elasticity of housing supply by assuming the log price of housing is governed by:  $r_{c} = k_{c} n_{c}$.  This is a reduced-form relationship between the log cost of housing and the log number of residents in city $c$.\footnote{The model assumes that housing is of constant quality, such that housing supply costs increase only with the number of residents.  Our focus is on changes in real housing costs, holding quality fixed, and in the empirical analysis we also present estimates that control for potential changes in housing quality.}  The parameter $k_{c}$ reflects differences in the elasticity of housing supply, which varies across cities due to differences in geographic constraints and local regulations on land development \cite{glaesergyourko2005}, \cite{glaeser2006urban, gyourko2009housing, saiz2010}.  

In cities where the geography and regulatory structure make it relatively easy to build new housing, $k_{c}$ is relatively smaller.  In the extreme case where there are no constraints on building housing, $k_{c}$ is zero and the supply curve is horizontal.  In the extreme case where it is impossible to build new housing, $k_{c}$ is infinite and the supply curve is vertical.\footnote{For simplicity, we are ignoring durability of the housing stock and the asymmetry between positive and negative shocks uncovered by \cite{glaesergyourko2005}.}

There are two types of residents. The first type receives a wage that is set competitively on the labor market.   Specifically, workers of this type  are paid their marginal product, and labor demand is derived from the usual first order conditions.\footnote{We abstract from labor supply decisions in the model and assume that each worker supplies one unit of labor.}  
The second type include workers who receives a wage that is not fully determined on the labor market.  
For simplicity, we will assume that workers of type 2 are paid a fixed wage. For example, they may be on welfare or on social security. Or they could be government employees, under the assumption that the government does not fully and immediately adjust wages after every shock. (In practice, of course, government do need to adjust wages to market conditions in the long run, but adjustment may not be immediate and even in the long run may not be complete.)

We are interested in what happens when one of the two cities  is hit by a positve shock to local rents. For example, consider the case where two cities are initially identical and  TFP increases in city $b$ by an amount $\Delta$. If $A_{b1}$ is initial TFP, the TFP gain is $A_{b2} - A_{b1} = \Delta$.  TFP in city $a$ does not change. 

Increased productivity in city $b$ shifts the local labor demand curve to the right, resulting in higher employment and higher nominal wages for type 1 workers.  Higher employment leads to higher housing costs. Assuming an interior solution, the changes in equilibrium of type 1 worker nominal wage and housing rent in city $b$ are:
\begin{eqnarray}
\label{eq_wage}
w_{b2} - w_{b1} = \frac{\beta (k_a + k_b) + h + s}{\beta (k_a + k_b) + 2 h  + s} \Delta > 0,
\end{eqnarray}
\begin{eqnarray}
\label{eq_rent}
r_{b2} - r_{b1} &= & \frac{ k_b }{\beta (k_a + k_b) + 2 h + s} \Delta > 0.
\end{eqnarray}
The magnitudes of these effects depend on the elasticities of labor supply and housing supply.   Nominal wages increase more when the elasticity of labor supply is lower ($s$ is larger), and housing costs increase more when the elasticity of housing supply in $b$ is lower ($k_b$ is larger). Type 2 worker wages are fixed by assumption, and therefore can't adjust to the shock.\footnote{To obtain equations \ref{eq_wage}, and \ref{eq_rent}, we equate local labor demand to local labor supply in each city and equate local housing demand to local housing supply in each city.  From the spatial equilibrium condition, the (inverse of) the \emph{local} labor supply to city $b$ in period $t$ is:  $w_{bt} =  w_{at}  + \beta (r_{bt} - r_{at}) + (x_{at} - x_{bt}) + 2 s (N_{bt} -1)$, where $N_{bt}$ is the share of employment in city $b$.  Since $N_{bt}$ is in levels, rather than logs, to obtain closed-form solutions in equations \ref{eq_wage}, and \ref{eq_rent}, we use a linear approximation around $1/2$: $n_{bt} = \ln N_{bt} \approx \ln(1/2) + 2 N_{bt} - 1$, so that we can assign $N_{bt} \approx (1/2) (n_{bt} - \ln(1/2) + 1)$ in the above equation for the (inverse of) the \emph{local} labor supply to city $b$ in period $t$.  We approximate around $1/2$ because of the assumption that the two cities are initially identical, which implies that their employment share is initially $1/2$.  We assume that local housing demand is proportional to city population.}

Following the shock, some workers move. First, as we mentioned, some type 1 workers move from city $a$ to city $b$, attracted by the stronger labor demand in $b$. Second, some type 2 workers leave city $b$ due to the higher cost of living and move to $a$, which is now more affordable. Since their wage is fixed, city $a$ affords them a higher level of non housing consumption.  
This process of mobility will continue until all workers have maximized their utility. 

Note that not all type $2$ workers abandon city $b$ to move to city $a$. Some decide to stay in $b$ because they have strong attachment to city $b$. In the model notation, the draw of  $e_{ic}$  makes their preferences for $b$ relative to $a$ particularly strong.  There is nothing in this model stopping them from relocating but these workers optimally choose $b$ over $a$ even if the cost of living in $b$ is higher, and their salary is fix. Clearly they are worse off after the TFP shock compared to before the shock, but even after the shock they are better off in $b$ than in $a$, otherwise they would have moved.

Turning to  consumption, the following expression describes the change of non-housing consumption of type 1 workers  in city $b$, defined as the increase in nominal wage minus the budget-share weighted increase in housing cost:
\begin{eqnarray} \label{eq_realwage}
(w_{b2} - w_{b1}) - \beta (r_{b2} - r_{b1}) = \frac{  \beta k_a + h + s}{\beta  (k_a + k_b) + 2h  + s} \Delta > 0.
\end{eqnarray}

Equation \ref{eq_realwage} shows how the benefits from productivity growth are split between workers and landowners, with the relative incidence depending on which of the two factors (labor or land) is supplied more elastically at the local level.  Intuitively, inelastically supplied factors should bear more incidence. For a given elasticity of housing supply, a lower local elasticity of labor supply (larger $s$) implies that a larger fraction of the productivity shock in city $b$ accrues to workers in city $b$ and that a smaller fraction accrues to landowners in city $b$.  When workers are less mobile, they capture more of the economic gains from local productivity growth.  In the extreme case, if labor is completely immobile ($s=\infty$), then equation \ref{eq_realwage} becomes:  $(w_{b2} - w_{b1}) - \beta(r_{b2} - r_{b1}) = \Delta$.  The real wage (or purchasing power) in city $b$ then increases by the full amount of the productivity shock, such that the benefit of the shock accrues entirely to workers in city $b$.  That is, when labor is a fixed factor, workers in the city directly impacted by the TFP shock will capture the full economic gain generated by the shock.

For a given elasticity of labor supply, a lower elasticity of housing supply in city $b$ (larger $k_b$) implies that a larger fraction of the productivity shock in city $b$ accrues to landowners in city $b$ and that a smaller fraction accrues to workers in city $b$.  When housing supply is more inelastic, the quantity of housing increases less in city $b$ and housing prices increase more following the local productivity shock.  In the extreme case, if housing supply in city $b$ is fixed ($k_b = \infty$), the entire productivity increase is capitalized into land values in city $b$ and worker purchasing power is unchanged.

Since their wage is fixed, the expression for the change of non-housing consumption of type 2 workers who stay in city $b$ is:
\begin{eqnarray} \label{eq_realwage2}
- \beta (\frac{ k_b }{\beta (k_a + k_b) + 2 h + s} \Delta ) < 0.
\end{eqnarray}

By contrast, the change of non-housing consumption of type 2 workers who move to city $a$ is positive, since their wage is unchanged and housing costs in $a$ have declined due to the out-migration to city $b$.

Motivated by equations \ref{eq_realwage} and \ref{eq_realwage2}, the empirical analysis will explore what happens to consumption of renters. We will first look at how the consumption of renters who are in a zip code that is hit by a sharp price increase changes in the years after the increase (compared to the consumption of renters who are in a zip code that is not hit by a sharp price increase changes in the years after the increase). We will then split the sample in two groups: Renters who relocate following the shock and renters who stay in the original zip code after the shock.  

In the context of our setting, the former group---made by those who leave---should include type 2 workers with weak preferences for their original location.  We expect them to move to areas that are cheaper and, as a consequence, to experience an increase in their non-housing consumption.  
The latter group---made of stayers---should include a combination of type 1 workers, whose wage adjusts to the shock and inframarginal type 2 workers, with preferences so strong that they are willing to remain in their original zip code despite the increase in rents. The effects on the average consumption of this group depends on whether the consumption increase experienced by  type 1 workers who stay is larger or smaller than the 
consumption decrease experienced by  type 2 workers who stay.

Empirically, the exact type of each worker is unknown. 


\section{Descriptive Statistics}
\label{sec:descriptives}

As mentioned above, our sample is drawn from the universe of individuals with a valid credit score in 2010. We then trace this population backwards and forward, at yearly frequency,  to 2004 and 2016 respectively. This gives a total of over 28 millions individual-year observations, and around 450 credit variables including basic demographics such as date of birth, zip code of residence, and two measures of income.

In the main text we use as definition of shock a sustained HPI cumulative increase of 25\% in the  previous 3 years up to 2007 or 2014.

\begin{figure}[ht!]
    \centering
    \begin{minipage}{\textwidth}
        \centering
        \includegraphics[width=0.9\textwidth]{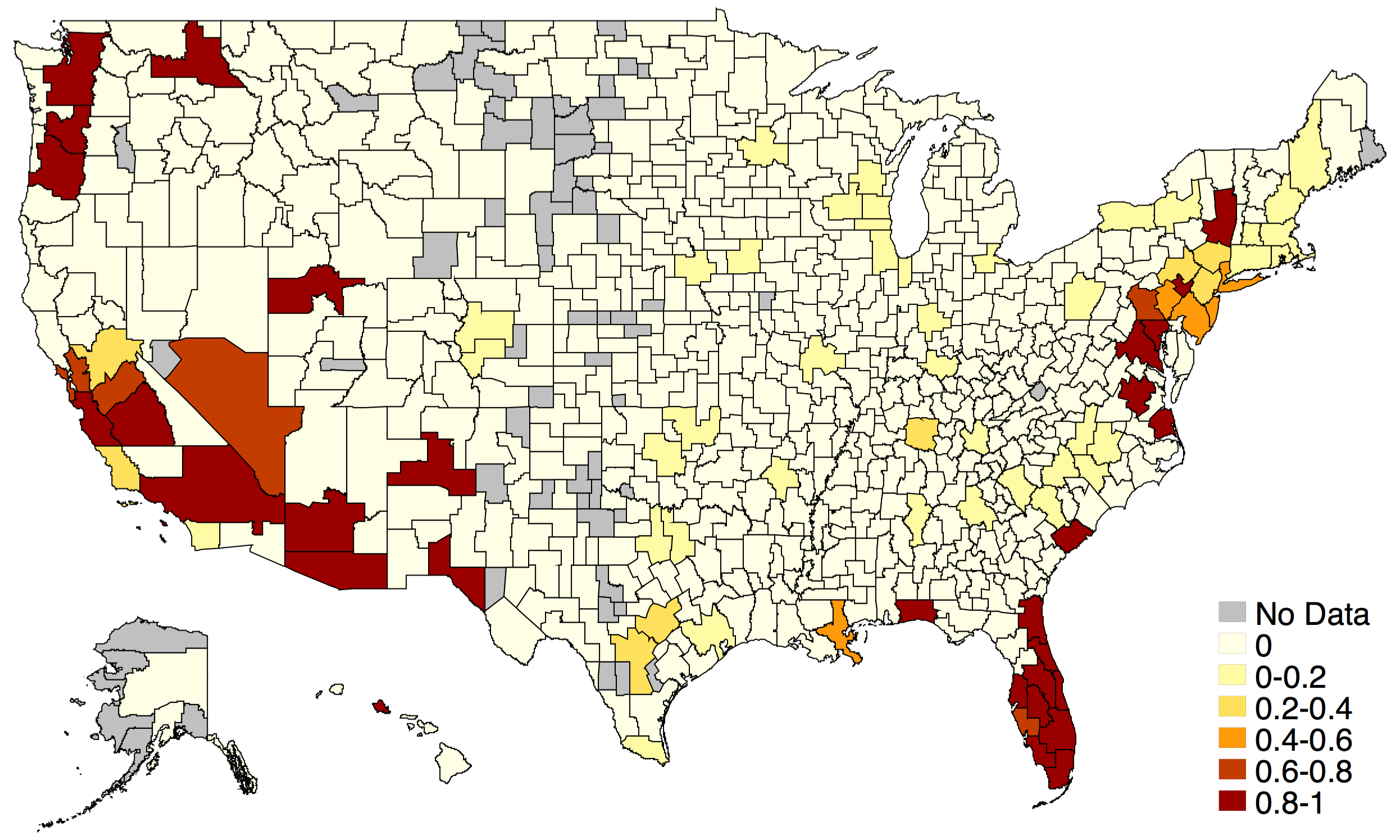}
        \caption*{Map of 2007 HPI Shocks (Share of CZ's ZIPs)}
        \label{fig:2007_shocks}
    \end{minipage}
    \hfill
    \begin{minipage}{\textwidth}
        \centering
        \includegraphics[width=0.9\textwidth]{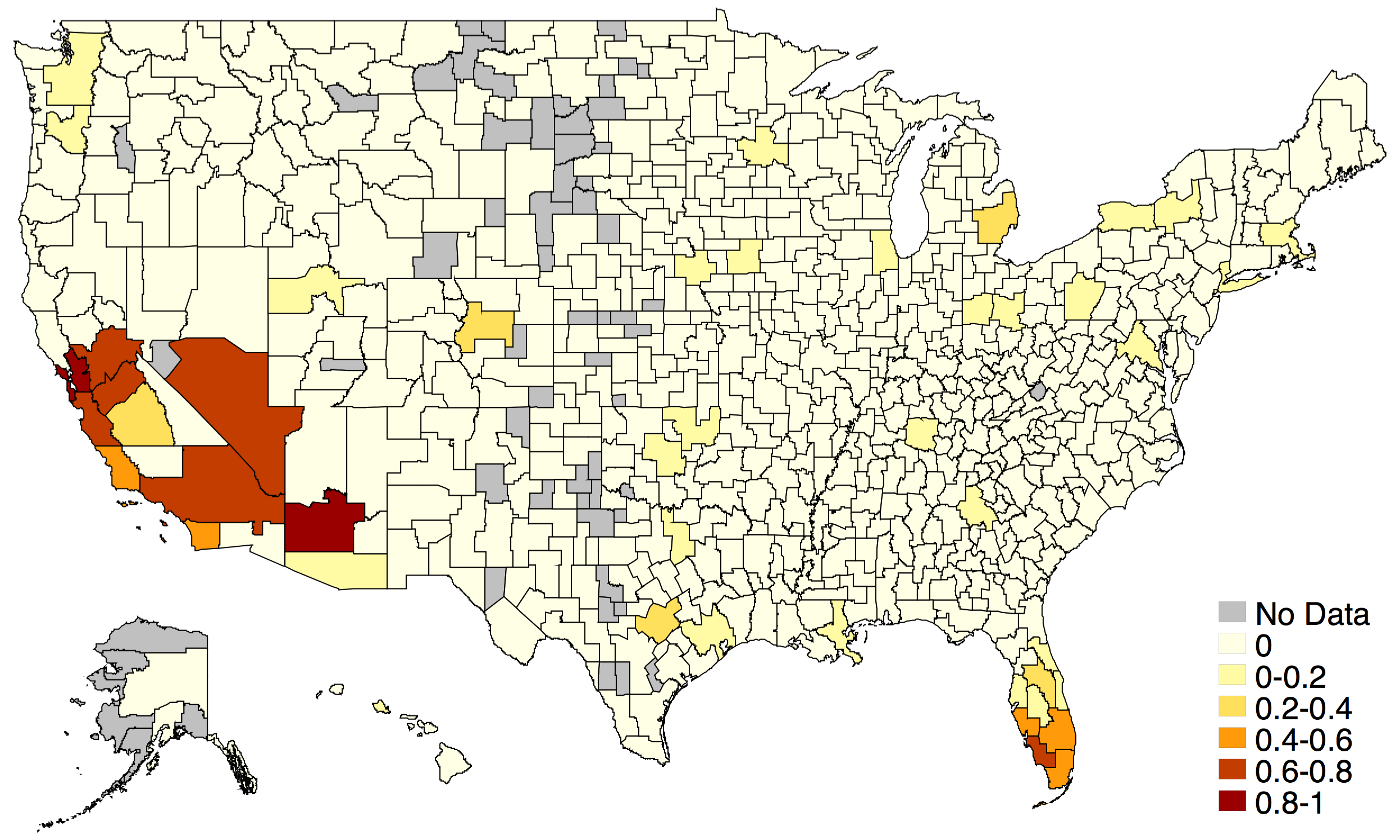}
        \caption*{Map of 2014 HPI Shocks (Share of ZIPs affected by CZs)}
        \label{fig:2014_shocks}
    \end{minipage}
    \caption{Comparison of HPI Shocks in 2007 and 2014}
    \label{fig:shocks_comparison}
\end{figure}

Figure \ref{fig:shocks_comparison} shows the areas with such a house price shock in terms of the shares of zip codes treated within the specific CZ for both the 2007 and the 2014 shock. It is easy to notice that the 2007 shocks affect mostly California, New York, and the sand States (Arizona, Nevada, Florida). While the 2014 shock appears more concentrated in California,  Nevada, and part of Florida.
However, we notice that there is substantial variation within CZs in Figure \ref{fig:cities}

Table \ref{tab:balance1} and \ref{tab:balance2} provide some basic description of the working sample of renters in treated and control areas as well as for movers and stayers.

A few facts stand out in the comparison between individuals in treated and control areas, at baseline: 1) treated areas comprise larger urban CZ; 2) younger individuals (41.6 vs. 44.3 years old on average); 3) lower average credit score (677 vs. 663); 4) about 5\% lower (yearly) income (38,140 vs. 36,350 USD); 5) slightly lower consumption (credit card balance) (3,855 vs 3,839 USD monthly); 6) similar mobility profiles before the shock zipcode-to-zipcode (0.13 vs. 0.16) or CZ-to-CZ (.05 vs. .05); 7) similar auto loan origination (.09) and with larger payments (4,336 vs. 4,616 USD monthly); 8) similar rate of Chapter 7 and Chapter 13 bankruptcy occurrences. 

We next characterize movers and stayers in treated areas. 
As expected movers are: 1) younger 39 vs 43 years of age; 2) have lower credit scores 647 vs. 674; 3) lower incomes 35,870 vs. 36,699 USD per year; 4) lower consumption; 5) the share of movers prior to the shock is substantially larger at 0.28 vs. 0.1 zip-to-zip, and 0.1 to 0.02 for CZ-to-CZ. This last fact is consistent with the theoretical framework, as movers are typically mobile people to start with so their attachment to a given location is not very high. Finally, we observe that movers have higher rates of auto loan origination (0.1 to 0.08) and larger monthly car payments (5,101 vs. 4,140 USD).

\section{Empirical Findings}
\label{sec:results}

\subsection{Event Study}
\label{sec:event_study}

We proceed as in a standard event study, where the estimating equation is:

\begin{equation}
y_{ict} = \alpha_i + \gamma_t + \sum_{j=-k,\neq-1}^{k} P_i \cdot 1(\tau=j) \cdot \beta_j + X_{it}'\zeta + \epsilon_{ict}
\end{equation}

with $y$ as the outcome of interest, i.e. mobility, consumption, mortgage origination, etc. for individual $i$, in CZ $c$, at time $t$. Because we have two periods of housing price shocks, and an individual could appear in neither, either, or both spells, $\alpha_i$ is an individual-spell fixed effect, where the housing price spell is either 2004-2010 or 2011-2016. The $\gamma_t$ are year fixed effects. $P_i$ is the event (price shock); and $\epsilon$  is an error term. The $X_{it}$ are controls we include to assess the robustness of our results to trends, e.g. commuting zone (at the time of the shock) by time fixed effects.

We are here interested  in the $\beta's$, which measure the effect of the HPI shock on our measures of consumption expenditures. 
The set of consumption expenditures, as dependent variables, comprises credit card balances (what we label as, consumption), mortgage balances, auto loans and in some cases mortgage originations, and auto loan originations.

In Table \ref{tab:HPI_Consumption}, we present the effects of the price shock for several measures of consumption and in particular we find that for our main measure of consumption (credit card balances), we have no pre-trends and that on average, for movers and stayers, consumption increases by 3 to 5\% in the 3 years following the shock. Recall that we cannot look beyond the 3 years period given the available data. We define the shocks as occurring in 2007 and 2014, so we only have 3 years before and after for non-overlapping periods. At the same time we see a fall in mortgage balances of about 2-6\%, lower originations of about 0.16-0.28pp (notice the mean of origination is 0.0128, so these are non-trivial magnitudes).
We also find a slight increase in auto loan balances, but no sizeable effects on originations. For this latter measure, however, it appears that the no pre-trends assumption is violated. 

\subsection{Movers and Stayers}

As described in Section \ref{sec:theory}, we believe that mobility is one of the main channels through which we should analyze the effects on consumption for renters. In these specifications, the dependent variable $y_{ict}$ is a dummy equal to 1 if the relevant consumer has changed zip code or CZ between $t-1$ and $t$. Alternatively,  $y_{ict}$ is a dummy equal to 1 if the relevant consumer has changed CZ between $t-1$ and $t$ and the destination zip code or CZ is 20\% less expensive than the origin CZ in terms of mean house price. 

What seems to happen is that people move, at least in the short-run to cheaper zip codes and CZs (see Table \ref{tab:HPI_Mobility}). These results are not overly sharp as we see some pre-trends and some lack of precision. 
 
The testable prediction is that individuals who move (to a cheaper location) will experience an increase in consumption due to lower housing costs, for a given income. For stayers, the net effect on consumption will depend on the income effects associated with the sharp increase in housing prices.  

\begin{figure}[ht!]
    \centering
    \begin{minipage}{0.45\textwidth}
        \centering
        \includegraphics[width=\textwidth]{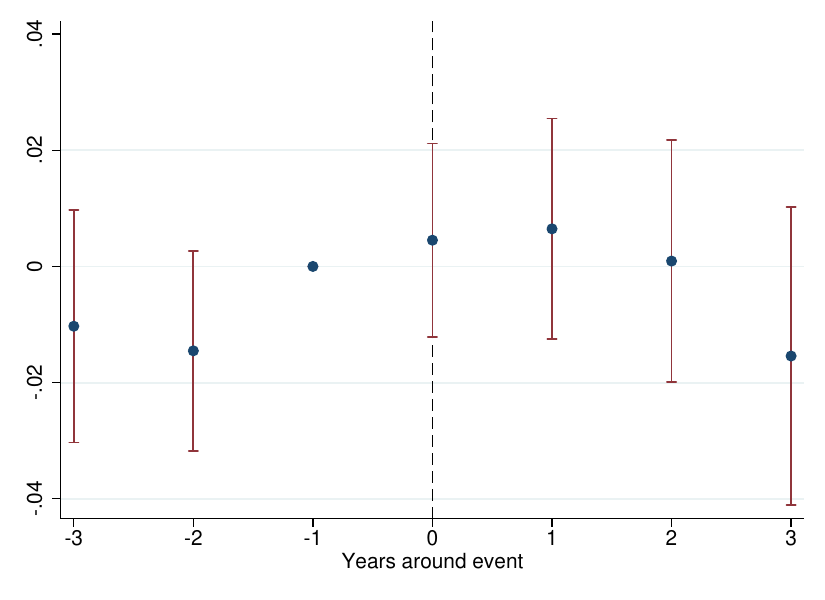}
        \caption*{Log Consumption - No Move}
        \label{fig:lcc_no_move}
    \end{minipage}
    \hfill
    \begin{minipage}{0.45\textwidth}
        \centering
        \includegraphics[width=\textwidth]{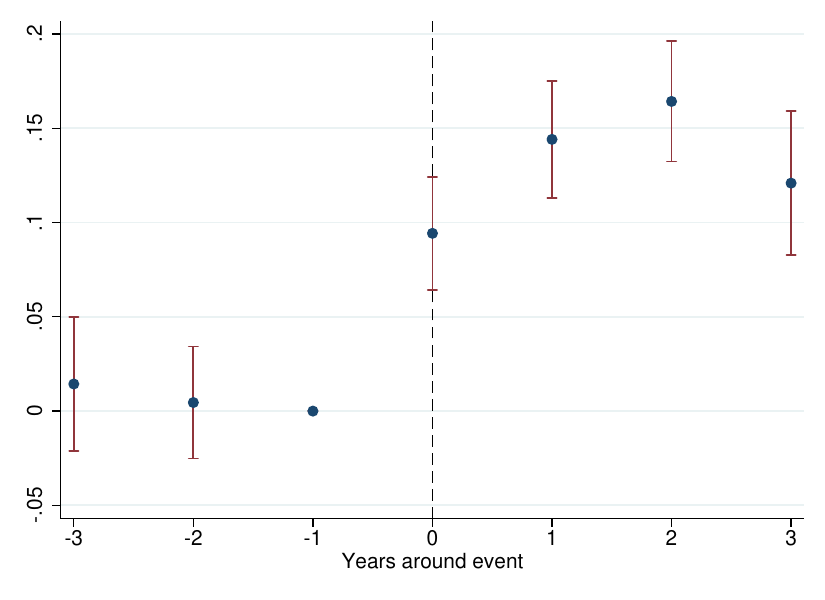}
        \caption*{Log Consumption - Move ($t=0$)}
        \label{fig:lcc_move_t0}
    \end{minipage}
    \caption{Effects on Log Consumption}
    \label{fig:log_consumption_comparison}
\end{figure}

\begin{figure}[ht!]
    \centering
    \begin{minipage}{0.45\textwidth}
        \centering
        \includegraphics[width=\textwidth]{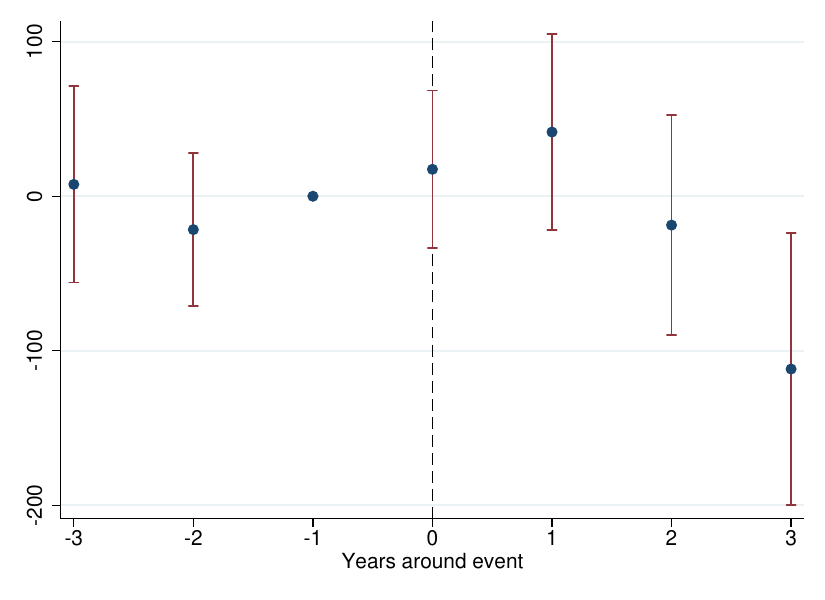}
        \caption*{Consumption - No Move}
        \label{fig:no_move_consumption}
    \end{minipage}
    \hfill
    \begin{minipage}{0.45\textwidth}
        \centering
        \includegraphics[width=\textwidth]{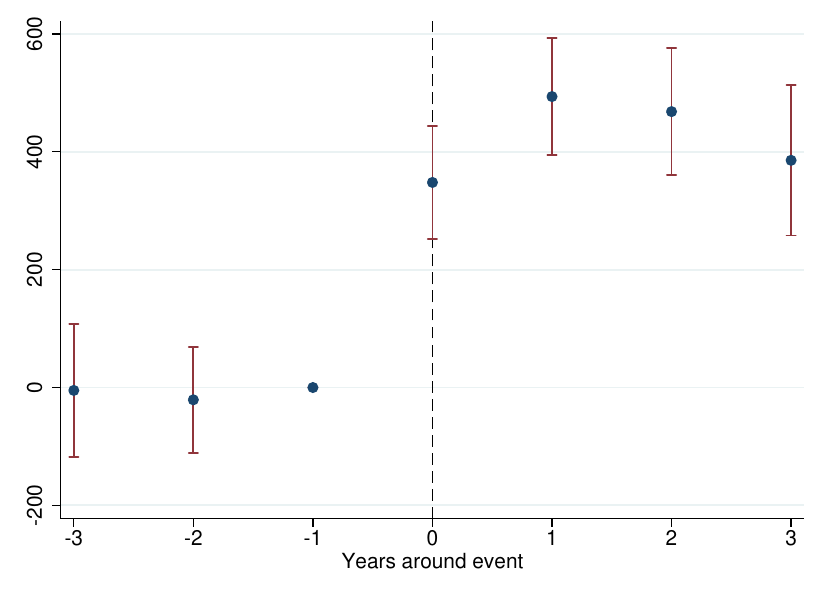}
        \caption*{Consumption - Move ($t=0$)}
        \label{fig:move_t0_consumption}
    \end{minipage}
    \caption{Effects on Consumption}
    \label{fig:consumption_comparison}
\end{figure}

We investigate exactly that in Figures \ref{fig:log_consumption_comparison} and \ref{fig:consumption_comparison}, the  coefficients, for movers, are in  Tables   \ref{tab:shock_stay} and \ref{tab:shock_move}. In Appendix Tables \ref{tab:table_full_nmv_match} and  \ref{tab:shock_move_match}, we provide a robustness check based on an adjusted propensity score method.

In the left panels of Figures \ref{fig:log_consumption_comparison} and \ref{fig:consumption_comparison} (the corresponding coefficients are in Table \ref{tab:shock_stay}) we study the impact on consumption for stayers and find that the impact is rather limited if any at all. 

In the right panels of Figures \ref{fig:log_consumption_comparison} and \ref{fig:consumption_comparison} (the corresponding coefficients are in Table \ref{tab:shock_move}) we study the impact on (log)consumption, and consumption, for movers and find that the impact is positive, large, and significant on impact and fairly stable for the next 3 years. 
In terms of credit card consumption, we see a pretty large increase for those who move, quantifiable in an increase on impact of about 9.5\% then rising to about 16.5\% by the second year, and remaining at a 12\% in the third year after the shock (if we use consumption levels, the effects are still sizeable and significant with magnitudes of 400-500 USD). 
Of course, mobility isn't exogenous, and that's what our theoretical framework suggests. 

\begin{figure}[ht!]
    \centering
    \begin{minipage}{0.45\textwidth}
        \centering
        \includegraphics[width=\textwidth]{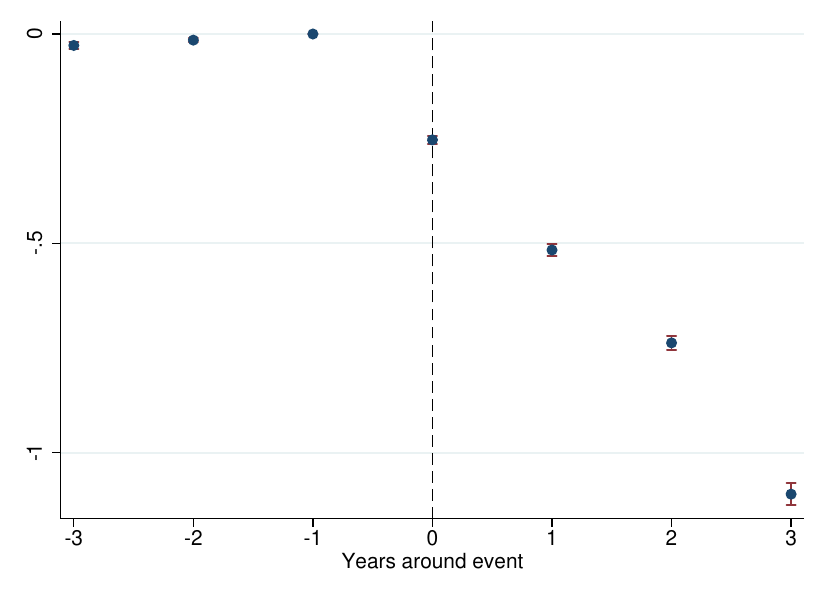}
        \caption*{Log Mortgage Balances - No Move}
        \label{fig:no_move_lmort}
    \end{minipage}
    \hfill
    \begin{minipage}{0.45\textwidth}
        \centering
        \includegraphics[width=\textwidth]{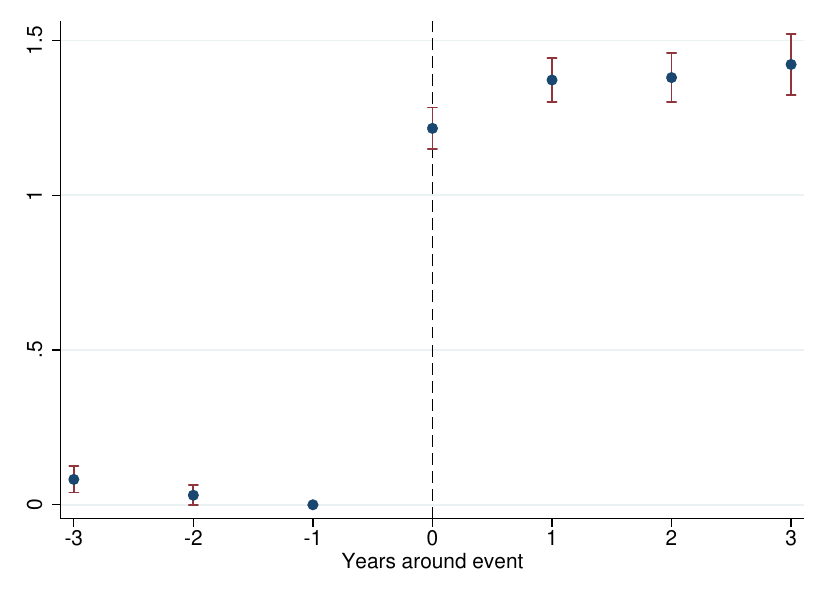}
        \caption*{Log Mortgage Balances - Move ($t=0$)}
        \label{fig:move_t0_lmort}
    \end{minipage}
    \caption{Effects on Log Mortgage Balances}
\label{fig:log_mortgage_balances_comparison}
\end{figure}

\begin{figure}[ht!]
    \centering
    \begin{minipage}{0.45\textwidth}
        \centering
        \includegraphics[width=\textwidth]{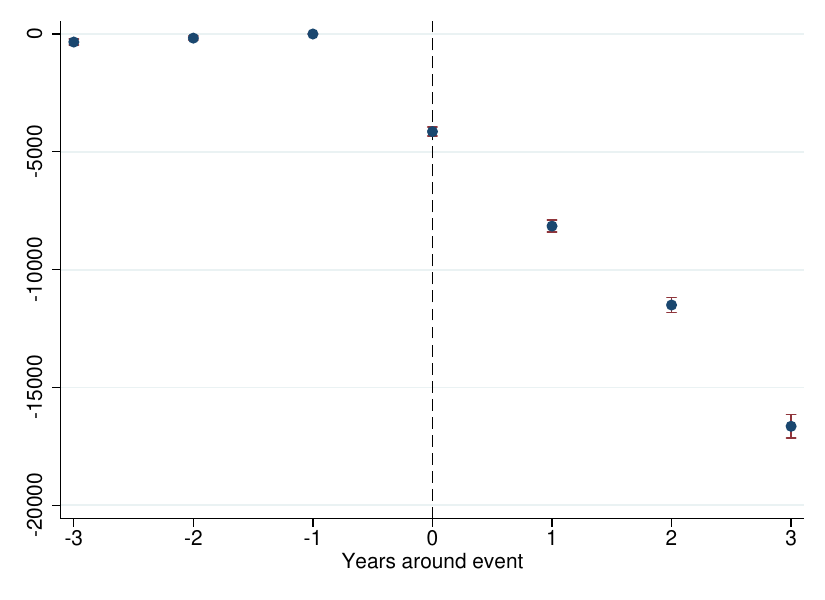}
        \caption*{Mortgage Balances - No Move}
        \label{fig:no_move_mortgage_balances}
    \end{minipage}
    \hfill
    \begin{minipage}{0.45\textwidth}
        \centering
        \includegraphics[width=\textwidth]{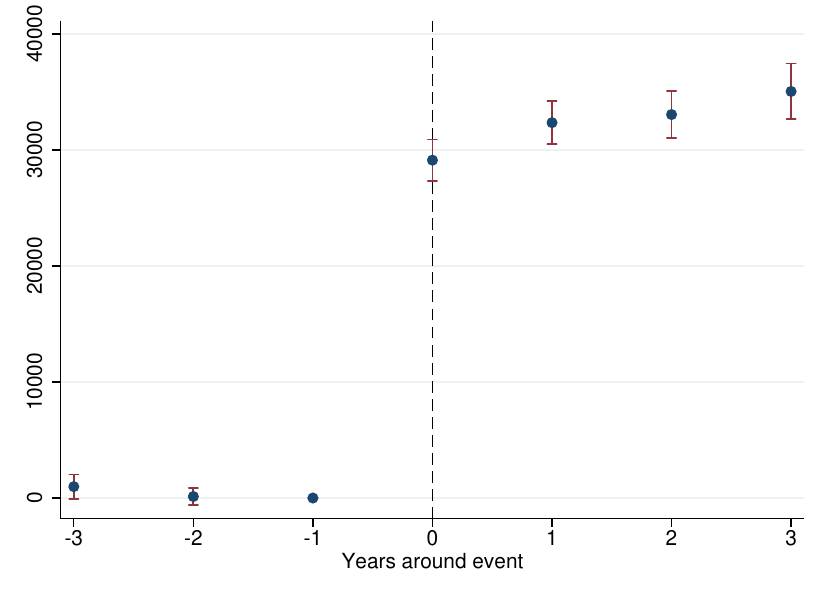}
        \caption*{Mortgage Balances - Move ($t=0$)}
        \label{fig:move_t0_mortgage_balances}
    \end{minipage}
    \caption{Effects on Mortgage Balances}
    \label{fig:mortgage_balances_comparison}
\end{figure}

\begin{figure}[ht!]
    \centering
    \begin{minipage}{0.45\textwidth}
        \centering
        \includegraphics[width=\textwidth]{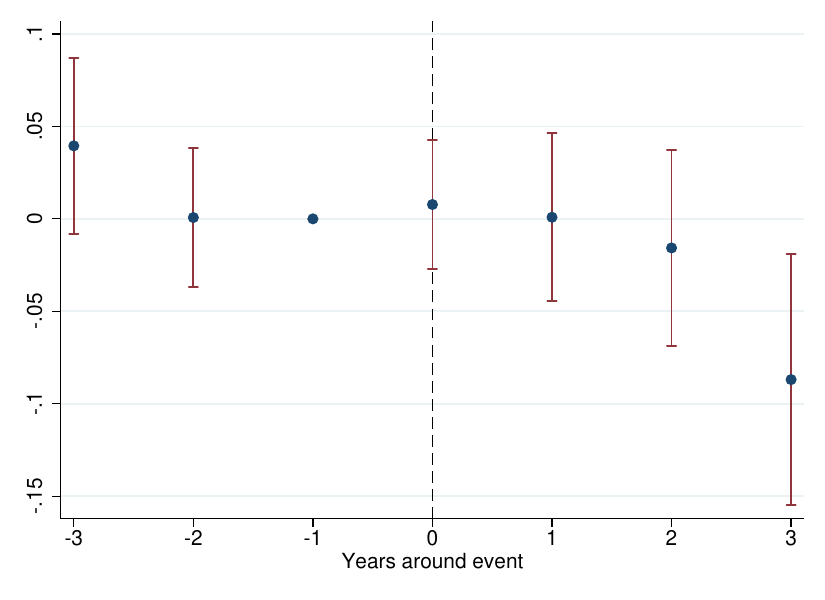}
        \caption*{Log Auto Balances - No Move}
    \label{fig:no_move_log_auto_balances}
    \end{minipage}
    \hfill
    \begin{minipage}{0.45\textwidth}
        \centering
        \includegraphics[width=\textwidth]{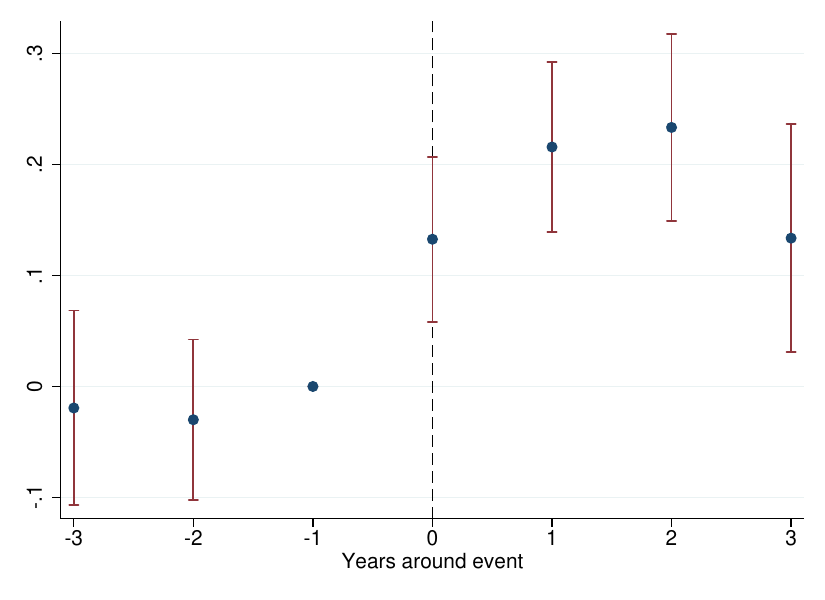}
        \caption*{Log Auto Balances - Move ($t=0$)}
        \label{fig:move_t0_log_auto_balances}
    \end{minipage}
    \caption{Effects on Log Auto Balances}
    \label{fig:log_auto_balances_comparison}
\end{figure}

\begin{figure}[ht!]
    \centering
    \begin{minipage}{.45\textwidth}
        \centering
        \includegraphics[width=\textwidth]{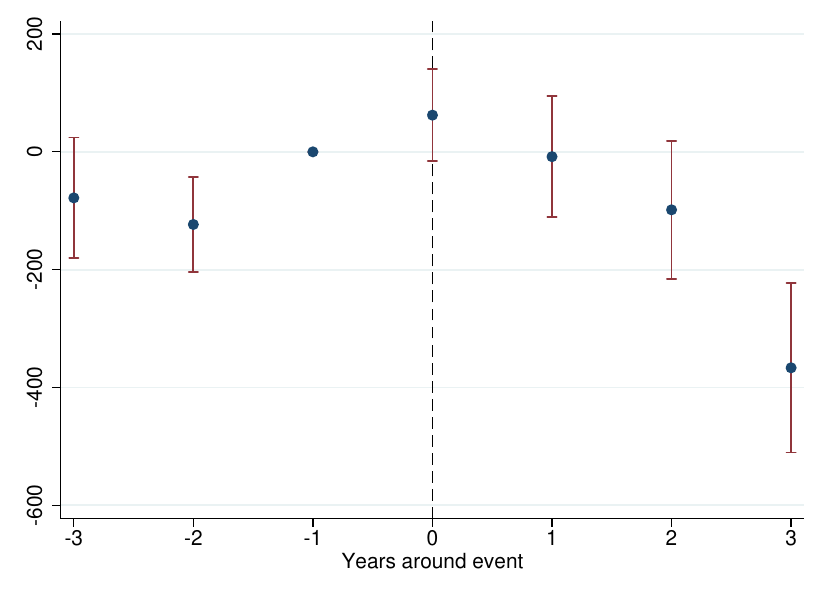}
        \caption*{No Move}
    \end{minipage}
    \hfill
    \begin{minipage}{.45\textwidth}
        \centering
        \includegraphics[width=\textwidth]{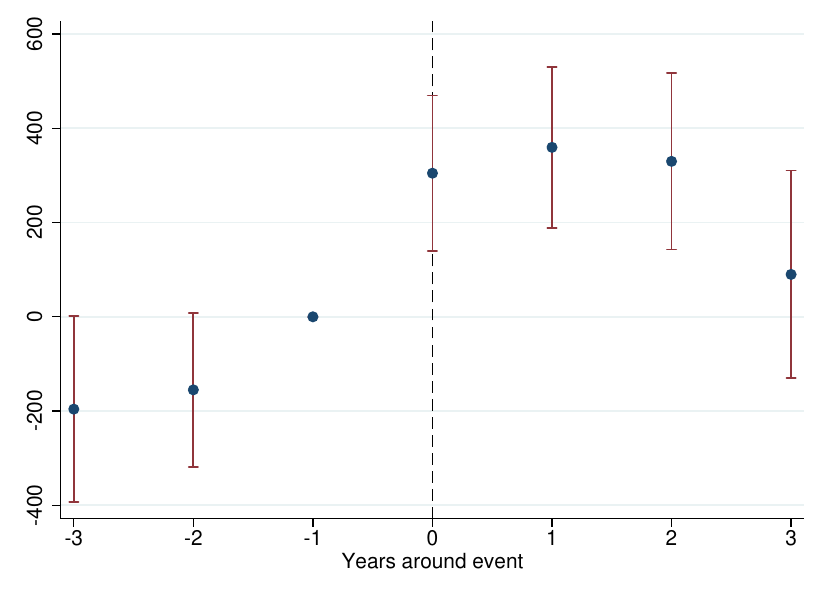}
        \caption*{Move ($t=0$)}
    \end{minipage}
    \caption{Effects on Auto Loan Balances }
    \label{fig:auto_balances_comparison}
\end{figure}

In terms of mortgage behavior documented in Figures \ref{fig:log_mortgage_balances_comparison} and \ref{fig:mortgage_balances_comparison}, we find a rather striking differential effect for movers versus stayers (irrespective of whether we use logs or levels). Essentially, people move to become homeowners. The effects on mortgage originations  are the largest on impact, which presumably coincides with the time of moving. 
 In terms of magnitudes, for stayers we see a fall in mortgage balances (or log) of 5,000 to 17,000 USD in the 3 years following the shock. These are falls of 20-100 log points. For movers, the story is totally different, mortgage balances go up by 110 to 150 log points, or in levels by 30-35 thousand USD. Mortgage origination are the essential part of the story with a fall of 2 percentage points for stayers and a jump up on impact of 9 percentage points then down to about 2 percentage points in the following years. The magnitudes on mortgage origination are very large, as the average origination is around 1.4pp. This suggests that upon moving, typically to a cheaper location, individuals open up new mortgages and keep larger balances. 
 Not only do movers buy properties, but they also seem to buy cars (Figures \ref{fig:log_auto_balances_comparison}, \ref{fig:auto_balances_comparison}, and Table \ref{tab:shock_move}). The auto loan balance increases 13.4\% on impact, and by 13.3\%, 21.6\%, and 23.3\% after 1, 2, and 3 years, respectively. On auto loan origination, we see a similar pattern with movers purchasing a new car on impact with a higher probability of 1.2 percentage points over a base of 8.4pp.

In Appendix \ref{sec:DID}, we confirm all the results running a more aggregate version of the previous analysis, where we condense the analysis in a simple two-periods differences-in-differences specification. That specification shows, in a compact manner, the dynamics presented in this section. As expected, the results are very much in line with those discussed above, if anything they are more precisely estimated.

 
 \section{Conclusions}
 \label{sec:conclusions}

In this paper we investigate, under the lenses of a spatial equilibrium model, the impact of sustained house prices increases, which we believe is part of what is commonly termed ``gentrification'', and how such increases impact individual
 consumption. 
 The narrative has often focused on homeowners and less is known for renters, who of course are different in many respects, but also because they generally do not see an increase in the value of their main assets, given that they do not own properties. 
 What we find is that renters' consumption remains relatively flat after such a shock, but this masks very important heterogeneity between those who stay and those who move. 
 Stayers, who probably benefit from the productivity increases which brought about the price shock, are not particularly affected as they are insured from the shock by capturing most of the productivity increases. Movers are rational, in the sense that they move to cheaper places. They might be in occupations without automatic productivity adjustments, and now they see an increase in their consumption -- a substantial one -- and they become homeowners as well as buy more and more expensive cars.

\newpage

\clearpage
\section{Bibliography}
\bibliography{biblio}
\clearpage

\newpage

\section{Figures}

  \clearpage
  \newpage

\clearpage


\clearpage

\vspace{30pt}

\newpage

 
  \begin{figure}[ht!]
    \centering
    \begin{minipage}{0.45\textwidth}
        \centering
        \includegraphics[width=\textwidth]{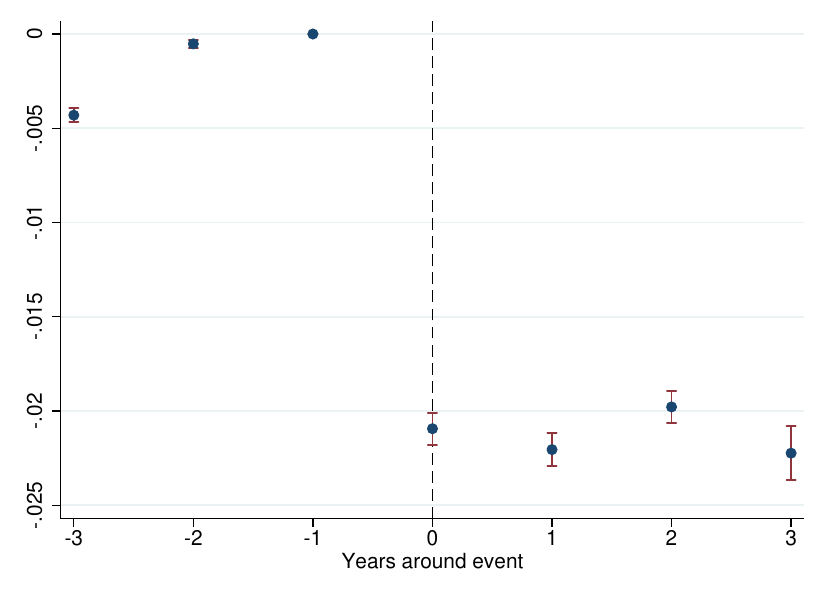}
        \caption*{Mortgage Origination - No Move}
        \label{fig:no_move_mortgage_origination}
    \end{minipage}
    \hfill
    \begin{minipage}{0.45\textwidth}
        \centering
        \includegraphics[width=\textwidth]{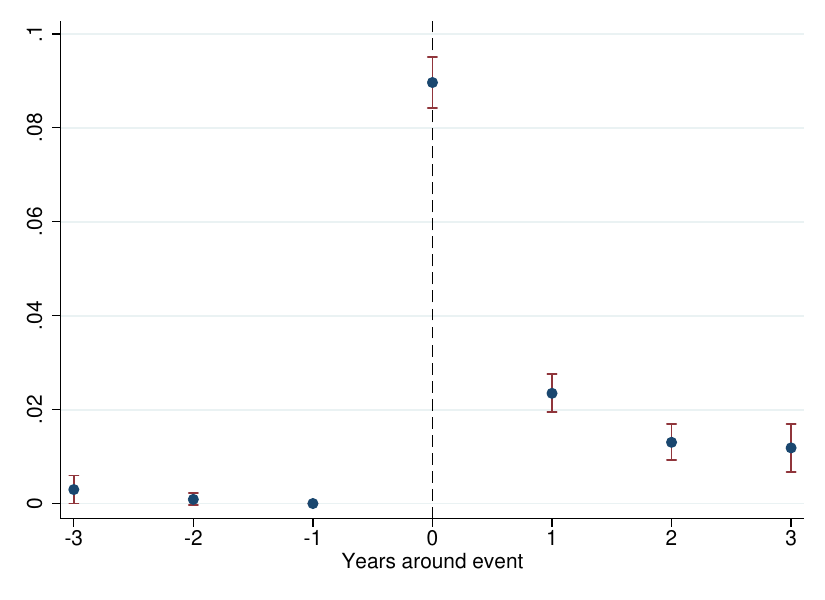}
        \caption*{Mortgage Origination - Move ($t=0$)}
        \label{fig:move_t0_mortgage_origination}
    \end{minipage}
    \caption{Effects on Mortgage Origination}
    \label{fig:mortgage_origination_comparison}
\end{figure}

  \pagebreak


 
\newpage


 
 \begin{figure}[ht!]
    \centering
    \begin{minipage}{.45\textwidth}
        \centering
        \includegraphics[width=\textwidth]{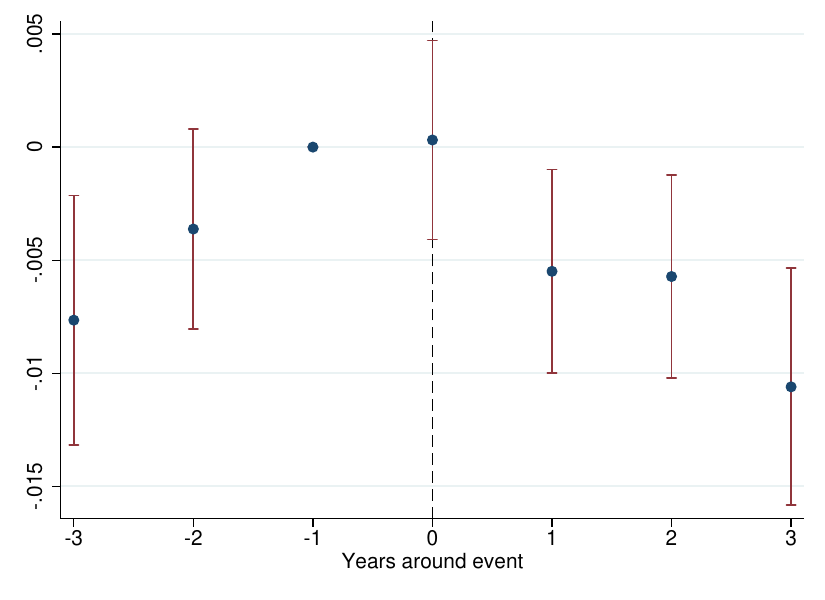}
        \caption*{No Move}
    \end{minipage}
    \hfill
    \begin{minipage}{.45\textwidth}
        \centering
        \includegraphics[width=\textwidth]{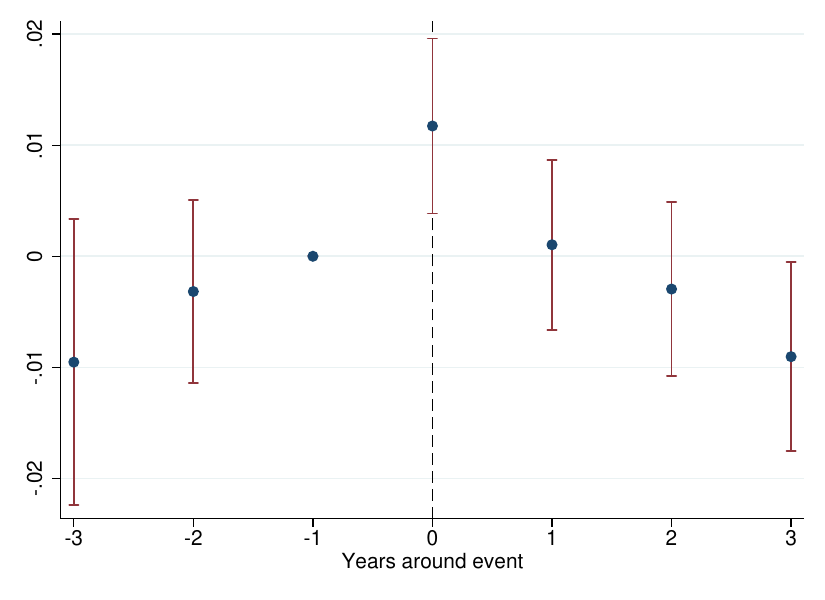}
        \caption*{Move ($t=0$)}
    \end{minipage}
    
    \caption{Effects on Auto Loan Originations}
    \label{fig:auto_originations_comparison}
\end{figure}

  \newpage
\pagebreak

  
\section{Tables}


    \begin{table}[ht!]
  \centering
  \resizebox{12cm}{!}{
\begin{tabular}{lccccc} \hline
 & (1) & (2) & (3) & (4) & (5) \\
VARIABLES & Log Consumption & Log Mortgage Bal. & Mortgage Orig. & Log Auto Bal. & Auto Orig. \\ \hline
\vspace{4pt} & \begin{footnotesize}\end{footnotesize} & \begin{footnotesize}\end{footnotesize} & \begin{footnotesize}\end{footnotesize} & \begin{footnotesize}\end{footnotesize} & \begin{footnotesize}\end{footnotesize} \\
3 yrs until shock & 0.00555 & 0.0135** & -0.000468 & 0.0226 & -0.00887*** \\
\vspace{4pt} & \begin{footnotesize}(0.00857)\end{footnotesize} & \begin{footnotesize}(0.00588)\end{footnotesize} & \begin{footnotesize}(0.000361)\end{footnotesize} & \begin{footnotesize}(0.0208)\end{footnotesize} & \begin{footnotesize}(0.00256)\end{footnotesize} \\
2 yrs until shock & -0.00391 & 0.00713 & 0.000284 & -0.0173 & -0.00469** \\
\vspace{4pt} & \begin{footnotesize}(0.00728)\end{footnotesize} & \begin{footnotesize}(0.00440)\end{footnotesize} & \begin{footnotesize}(0.000181)\end{footnotesize} & \begin{footnotesize}(0.0165)\end{footnotesize} & \begin{footnotesize}(0.00195)\end{footnotesize} \\
shock & 0.0207*** & -0.0192** & -0.00227*** & 0.0298* & 0.000928 \\
\end{tabular}
}
  \captionof{table}{HPI Shock on Consumption}
\label{tab:HPI_Consumption}
   \end{table}

\newpage
    \begin{table}[ht!]
  \centering
  \resizebox{12cm}{!}{
\begin{tabular}{lccccc} \hline
 & (1) & (2) & (3) & (4) & (5) \\
VARIABLES & Move Zip & Move CZ & Move Zip (cheaper) & Move CZ (cheaper zip) & Move CZ (cheaper) \\ \hline
\vspace{4pt} & \begin{footnotesize}\end{footnotesize} & \begin{footnotesize}\end{footnotesize} & \begin{footnotesize}\end{footnotesize} & \begin{footnotesize}\end{footnotesize} & \begin{footnotesize}\end{footnotesize} \\
3 yrs until shock & 0.00887*** & -0.000613 & 0.0117*** & 0.00724*** & 0.00813*** \\
\vspace{4pt} & \begin{footnotesize}(0.00340)\end{footnotesize} & \begin{footnotesize}(0.00182)\end{footnotesize} & \begin{footnotesize}(0.00263)\end{footnotesize} & \begin{footnotesize}(0.00130)\end{footnotesize} & \begin{footnotesize}(0.00136)\end{footnotesize} \\
2 yrs until shock & 0.00471** & -0.000932 & 0.00387** & 0.00156 & 0.00282*** \\
\vspace{4pt} & \begin{footnotesize}(0.00231)\end{footnotesize} & \begin{footnotesize}(0.00138)\end{footnotesize} & \begin{footnotesize}(0.00184)\end{footnotesize} & \begin{footnotesize}(0.000977)\end{footnotesize} & \begin{footnotesize}(0.00102)\end{footnotesize} \\
shock & -0.000976 & -0.00243 & 0.0145*** & 0.0151*** & 0.0198*** \\
\end{tabular}
  }
  \captionof{table}{HPI Shock on Mobility}
\label{tab:HPI_Mobility}
   \end{table}

\newpage

\begin{table}[ht!]
    \centering
    \resizebox{12cm}{!}{
\begin{tabular}{lccccc} \hline
 & (1) & (2) & (3) & (4) & (5) \\
VARIABLES & Log Consumption & Log Mortgage Bal. & Mortgage Orig. & Log Auto Bal. & Auto Orig. \\ \hline
\vspace{4pt} & \begin{footnotesize}\end{footnotesize} & \begin{footnotesize}\end{footnotesize} & \begin{footnotesize}\end{footnotesize} & \begin{footnotesize}\end{footnotesize} & \begin{footnotesize}\end{footnotesize} \\
3 yrs until shock & -0.0103 & -0.0273*** & -0.00431*** & 0.0395 & -0.00765*** \\
\vspace{4pt} & \begin{footnotesize}(0.0102)\end{footnotesize} & \begin{footnotesize}(0.00385)\end{footnotesize} & \begin{footnotesize}(0.000193)\end{footnotesize} & \begin{footnotesize}(0.0243)\end{footnotesize} & \begin{footnotesize}(0.00281)\end{footnotesize} \\
2 yrs until shock & -0.0145* & -0.0147*** & -0.000522*** & 0.000692 & -0.00362 \\
\vspace{4pt} & \begin{footnotesize}(0.00878)\end{footnotesize} & \begin{footnotesize}(0.00275)\end{footnotesize} & \begin{footnotesize}(0.000109)\end{footnotesize} & \begin{footnotesize}(0.0191)\end{footnotesize} & \begin{footnotesize}(0.00226)\end{footnotesize} \\
shock & 0.00452 & -0.253*** & -0.0209*** & 0.00776 & 0.000314 \\
\end{tabular}
}
        \captionof{table}{HPI Shock on Consumption Dynamics for Stayers }
        \label{tab:shock_stay}
\end{table}

\begin{table}[ht!]
    \centering
    \resizebox{12cm}{!}{
\begin{tabular}{lccccc} \hline
 & (1) & (2) & (3) & (4) & (5) \\
VARIABLES & Log Consumption & Log Mortgage Bal. & Mortgage Orig. & Log Auto Bal. & Auto Orig. \\ \hline
\vspace{4pt} & \begin{footnotesize}\end{footnotesize} & \begin{footnotesize}\end{footnotesize} & \begin{footnotesize}\end{footnotesize} & \begin{footnotesize}\end{footnotesize} & \begin{footnotesize}\end{footnotesize} \\
3 yrs until shock & 0.0144 & 0.0819*** & 0.00298* & -0.0194 & -0.00953 \\
\vspace{4pt} & \begin{footnotesize}(0.0182)\end{footnotesize} & \begin{footnotesize}(0.0217)\end{footnotesize} & \begin{footnotesize}(0.00152)\end{footnotesize} & \begin{footnotesize}(0.0448)\end{footnotesize} & \begin{footnotesize}(0.00657)\end{footnotesize} \\
2 yrs until shock & 0.00453 & 0.0308* & 0.000884 & -0.0300 & -0.00317 \\
\vspace{4pt} & \begin{footnotesize}(0.0152)\end{footnotesize} & \begin{footnotesize}(0.0167)\end{footnotesize} & \begin{footnotesize}(0.000650)\end{footnotesize} & \begin{footnotesize}(0.0369)\end{footnotesize} & \begin{footnotesize}(0.00419)\end{footnotesize} \\
shock & 0.0943*** & 1.216*** & 0.0897*** & 0.133*** & 0.0117*** \\
\end{tabular}
}
        \captionof{table}{HPI Shock on Consumption Dynamics for Movers  ($t=0$)}
        \label{tab:shock_move}
\end{table}


\clearpage

\appendix

\noindent\section{Figures and Tables}

\renewcommand{\thefigure}{A.\arabic{figure}}
\renewcommand{\thetable}{A.\arabic{table}}

\setcounter{figure}{0}
\setcounter{table}{0}

\subsection{Figures}

    \begin{figure}[h!]
      \centering
    \includegraphics[scale=.7]{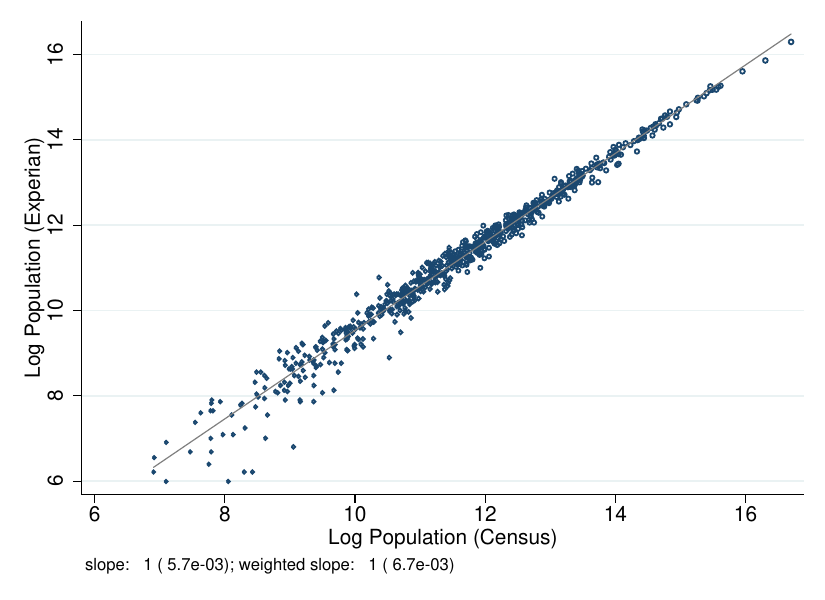}
    \caption{CZ Population (vs. Census)}
   \label{fig:a1}
    \end{figure}

      \begin{figure}
       \centering
    \includegraphics[scale=.7]{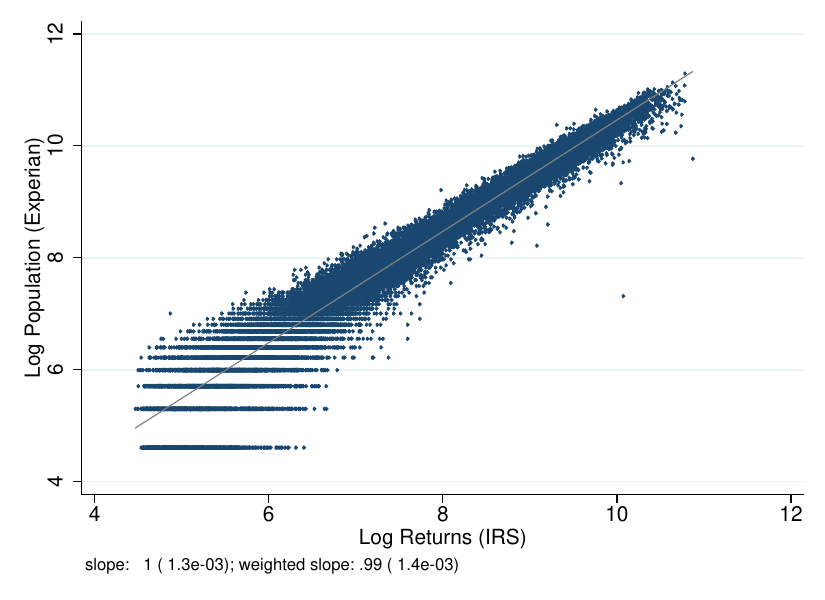}
   \caption{ZIP Code Population (vs. IRS returns)}
       \label{fig:a2}
    \end{figure}

\newpage
      \begin{figure}
    \centering
    \includegraphics[scale=.7]{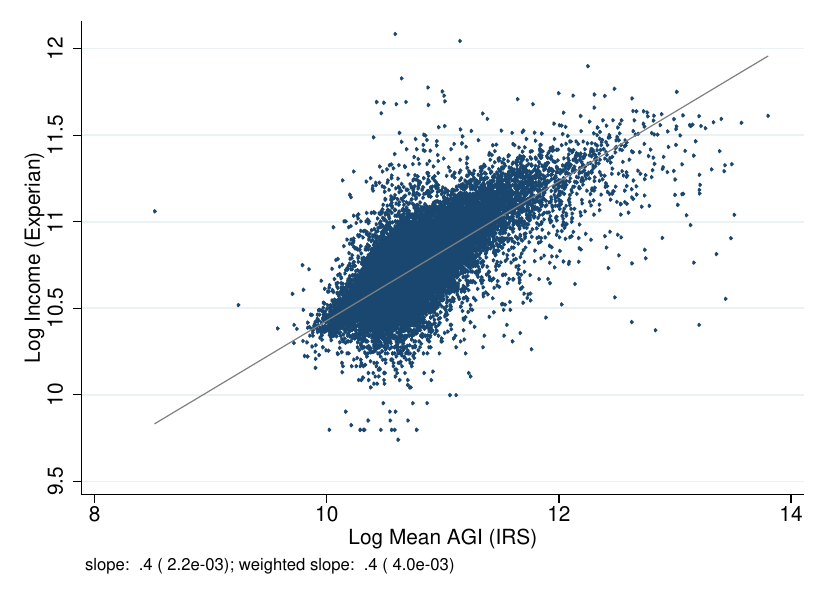}
    \caption{ZIP Code Income (vs. IRS)}
    \label{fig:a3}
    \end{figure}

 \begin{figure}
    \centering
    \includegraphics[scale=.7]{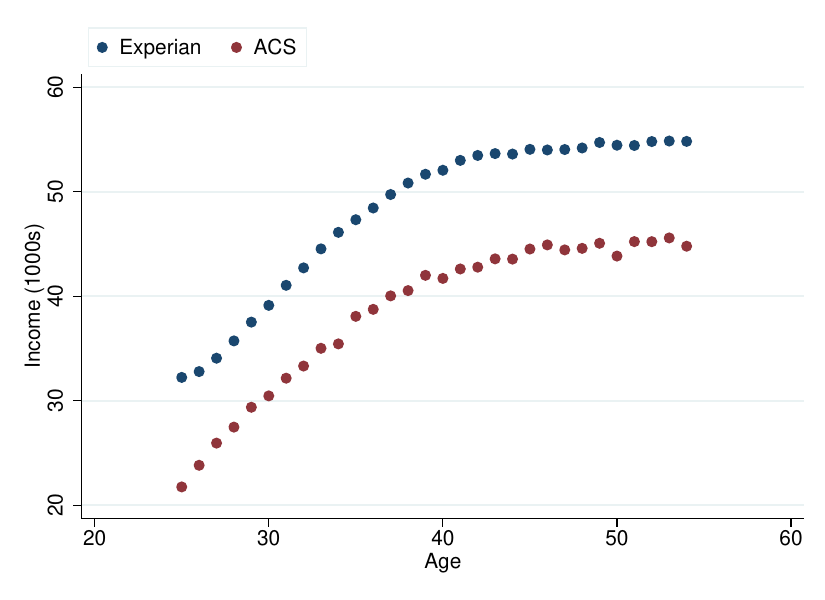}
    \caption{Lifecycle Income (vs. ACS)}
    \label{fig:a4}
    \end{figure}

  \newpage

      \begin{figure}
    \centering
    \includegraphics[scale=.7]{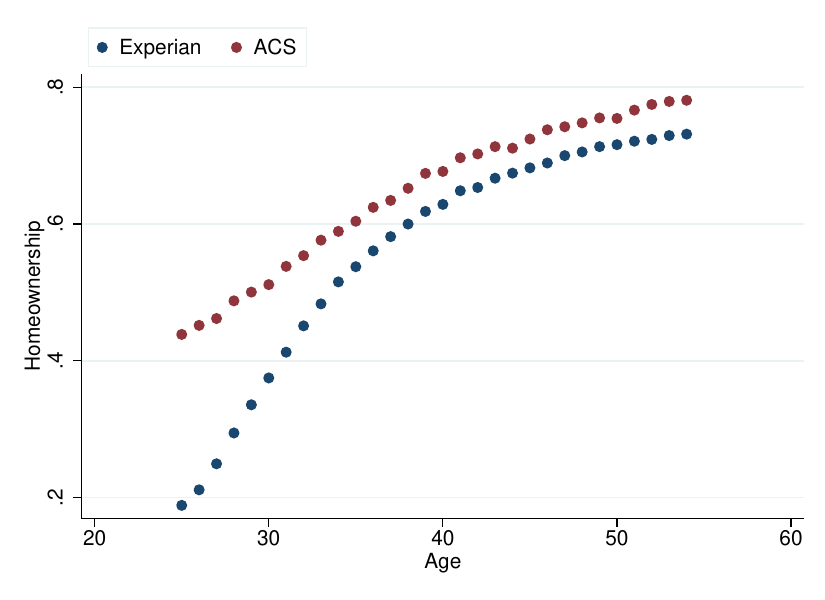}
    \caption{Lifecycle Homeownership}
    \label{fig:a6}
    \end{figure}
    
  \newpage
  
      \begin{figure}
    \centering
    \includegraphics[scale=.7]{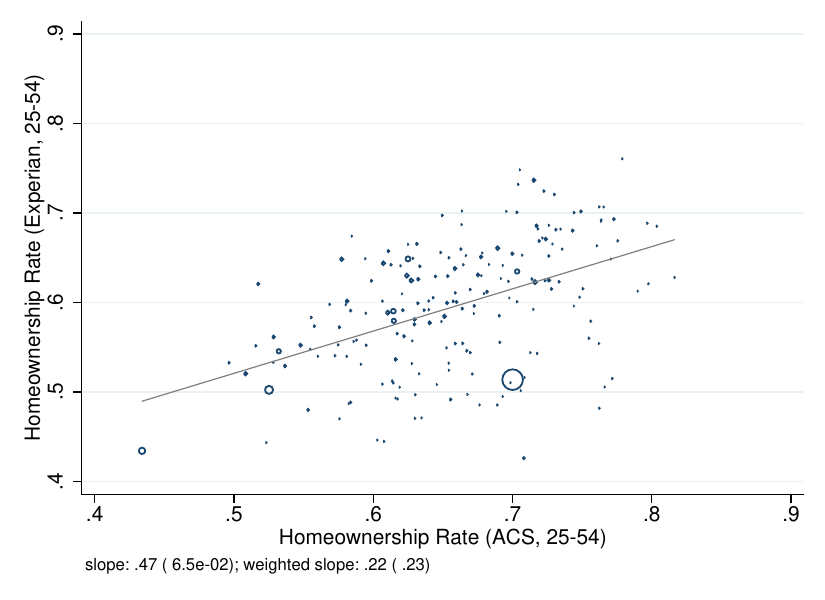}
    \caption{CZ Homeownership}
    \label{fig:a7}
    \end{figure}

 \newpage
  
      \begin{figure}
    \centering
    \includegraphics[scale=.7]{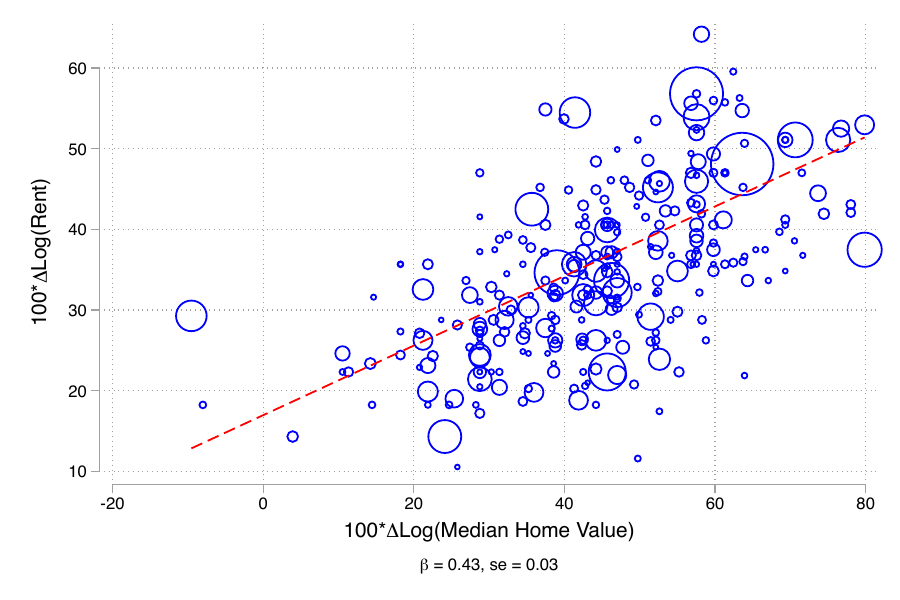}
    \caption{Rents and home values 2000-2010}
    \label{fig:a_rent}
    \end{figure}

\clearpage
   
  \begin{figure}[ht!]
    \begin{minipage}[b]{0.45\linewidth}
      \centering
      \includegraphics[width=\textwidth]{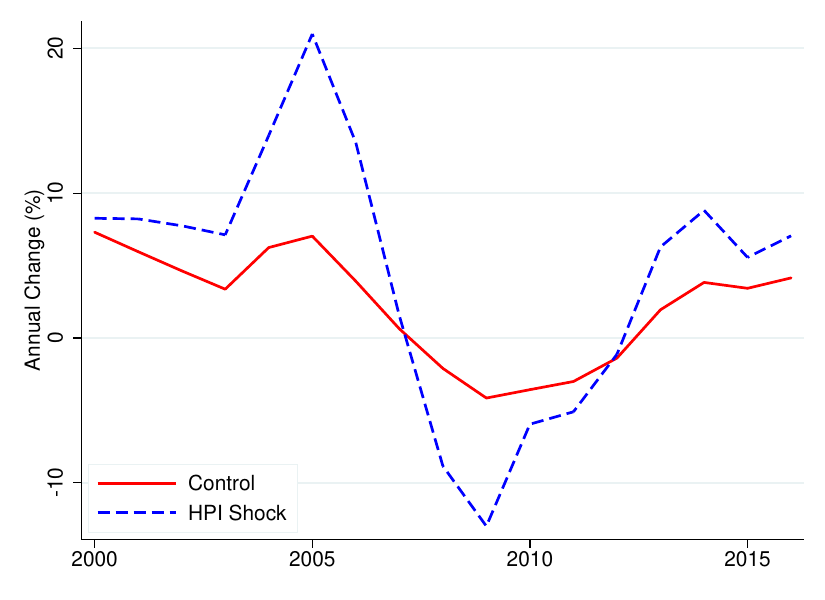}
      \vspace{\abovecaptionskip}
      Growth
    \end{minipage}
    \begin{minipage}[b]{0.45\linewidth}
      \centering
      \includegraphics[width=\textwidth]{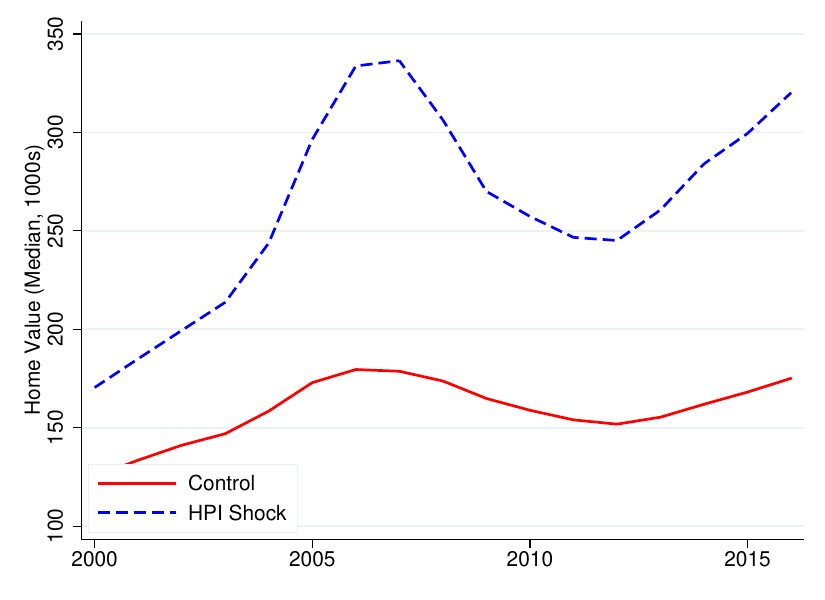}
      \vspace{\abovecaptionskip}
      Home Value
    \end{minipage}
    \caption{2007 Price Shock}
    \label{fig:a8}
  \end{figure}

  \begin{figure}[ht!]
    \begin{minipage}[b]{0.45\linewidth}
      \centering
      \includegraphics[width=\textwidth]{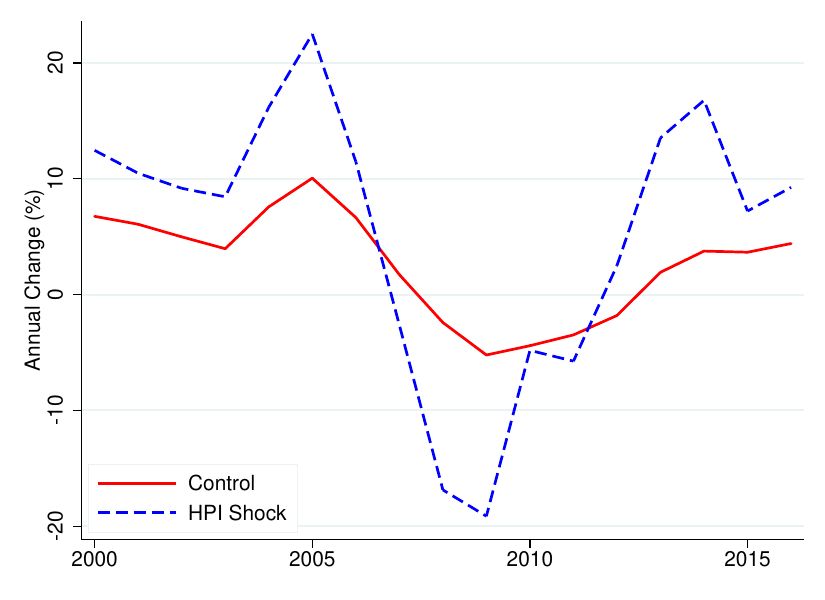}
      \vspace{\abovecaptionskip}
      Growth
    \end{minipage}
    \begin{minipage}[b]{0.45\linewidth}
      \centering
      \includegraphics[width=\textwidth]{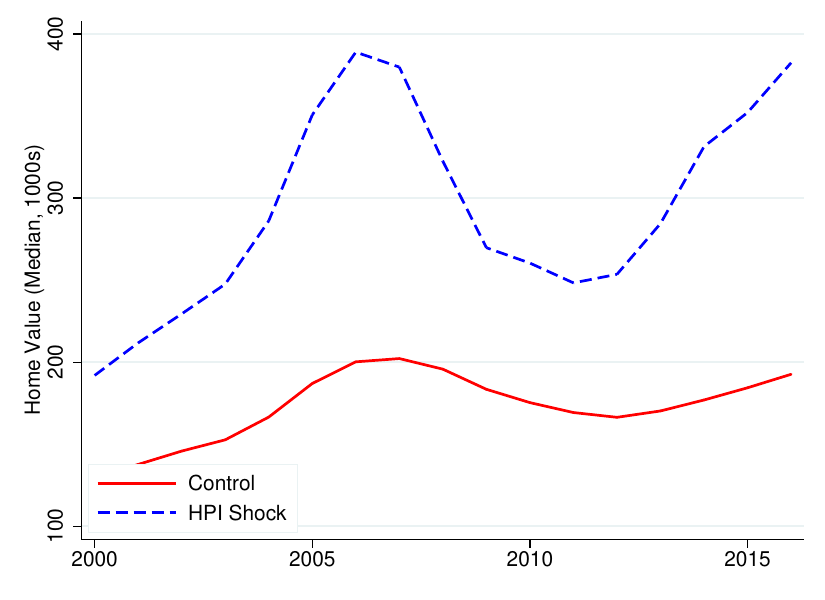}
      \vspace{\abovecaptionskip}
      Home Value
    \end{minipage}
     \caption{2014 Price Shock}
     \label{fig:a9}
  \end{figure}

\newpage
\subsection{Tables}
  
  \begin{table}[ht!]
    \centering
    \resizebox{10cm}{!}{%
        {
\def\sym#1{\ifmmode^{#1}\else\(^{#1}\)\fi}
\begin{tabular}{l*{5}{cc}}
\toprule
                    &\multicolumn{1}{c}{(1)}&\multicolumn{1}{c}{(2)}&\multicolumn{1}{c}{(3)}&\multicolumn{1}{c}{(4)}&\multicolumn{1}{c}{(5)}\\
                    &\multicolumn{1}{c}{Total}&\multicolumn{1}{c}{Control}&\multicolumn{1}{c}{Treated}&\multicolumn{1}{c}{Diff.}&\multicolumn{1}{c}{Diff. (w/ FE)}\\
                    &        mean&        mean&        mean&           b         &           b         \\
\midrule
Age                 &       43.46&       44.34&       41.56&        2.78\sym{***}&        0.00\sym{***}\\
Credit Score        &      673.01&      677.48&      663.40&       14.08\sym{***}&       -0.49\sym{**} \\
Income              &    37571.21&    38139.71&    36349.61&     1790.10\sym{***}&      256.91\sym{***}\\
Consumption         &     3849.76&     3854.81&     3838.91&       15.91         &       56.80\sym{***}\\
Move Zip            &        0.14&        0.13&        0.16&       -0.04\sym{***}&        0.00         \\
Move CZ             &        0.05&        0.05&        0.05&       -0.00\sym{*}  &       -0.00         \\
Auto Orig.          &        0.09&        0.09&        0.09&       -0.01\sym{***}&       -0.00\sym{**} \\
Auto Bal.           &     4424.75&     4335.97&     4615.64&     -279.67\sym{***}&      -40.70         \\
\midrule
Observations        &      251573&      171710&       79863&      251573         &      246801         \\
\bottomrule
\end{tabular}
}
    }
     \captionof{table}{Summary Statistics for $t=-1$: Treat vs. Control}
    \label{tab:balance1}
\end{table}

    \begin{table}[ht!]
  \centering
  \resizebox{10cm}{!}{
    {
\def\sym#1{\ifmmode^{#1}\else\(^{#1}\)\fi}
\begin{tabular}{l*{5}{cc}}
\toprule
                    &\multicolumn{1}{c}{(1)}&\multicolumn{1}{c}{(2)}&\multicolumn{1}{c}{(3)}&\multicolumn{1}{c}{(4)}&\multicolumn{1}{c}{(5)}\\
                    &\multicolumn{1}{c}{Treated}&\multicolumn{1}{c}{No Move}&\multicolumn{1}{c}{Move (t=0)}&\multicolumn{1}{c}{Diff.}&\multicolumn{1}{c}{Diff. (w/ FE)}\\
                    &        mean&        mean&        mean&           b         &           b         \\
\midrule
Age                 &       41.56&       43.32&       38.90&        4.42\sym{***}&        0.00\sym{***}\\
Credit Score        &      663.40&      674.06&      647.49&       26.57\sym{***}&       -2.34\sym{***}\\
Income              &    36349.61&    36698.63&    35869.69&      828.94\sym{***}&     1701.70\sym{***}\\
Consumption         &     3838.91&     3878.88&     3732.28&      146.60\sym{**} &      221.09\sym{***}\\
Move Zip            &        0.16&        0.10&        0.28&       -0.18\sym{***}&        0.17\sym{***}\\
Move CZ             &        0.05&        0.02&        0.10&       -0.07\sym{***}&        0.03\sym{***}\\
Auto Orig.          &        0.09&        0.08&        0.10&       -0.02\sym{***}&        0.00         \\
Auto Bal.           &     4615.64&     4140.44&     5100.99&     -960.55\sym{***}&      174.16\sym{***}\\
\midrule
Observations        &       79863&       46940&       13825&       60765         &       59693         \\
\bottomrule
\end{tabular}

}
  }
  \captionof{table}{Summary Statistics for $t=-1$: Move vs. No Move}
    \label{tab:balance2}
  \end{table}

  \newpage

\newpage
\begin{table}[ht!]
    \centering
    \resizebox{12cm}{!}{
\begin{tabular}{lccccc} \hline
 & (1) & (2) & (3) & (4) & (5) \\
VARIABLES & Log Consumption & Log Mortgage Bal. & Mortgage Orig. & Log Auto Bal. & Auto Orig. \\ \hline
\vspace{4pt} & \begin{footnotesize}\end{footnotesize} & \begin{footnotesize}\end{footnotesize} & \begin{footnotesize}\end{footnotesize} & \begin{footnotesize}\end{footnotesize} & \begin{footnotesize}\end{footnotesize} \\
3 yrs until shock & -0.00701 & -0.0291*** & -0.00408*** & 0.0943** & 0.00412 \\
\vspace{4pt} & \begin{footnotesize}(0.0203)\end{footnotesize} & \begin{footnotesize}(0.00881)\end{footnotesize} & \begin{footnotesize}(0.000528)\end{footnotesize} & \begin{footnotesize}(0.0465)\end{footnotesize} & \begin{footnotesize}(0.00520)\end{footnotesize} \\
2 yrs until shock & 0.00487 & -0.0141** & -0.000531* & 0.0286 & -0.00258 \\
\vspace{4pt} & \begin{footnotesize}(0.0167)\end{footnotesize} & \begin{footnotesize}(0.00661)\end{footnotesize} & \begin{footnotesize}(0.000295)\end{footnotesize} & \begin{footnotesize}(0.0367)\end{footnotesize} & \begin{footnotesize}(0.00460)\end{footnotesize} \\
shock & -0.0279* & -0.265*** & -0.0219*** & -0.0484 & -0.000342 \\
\end{tabular}
}
        \captionof{table}{HPI Shock on Consumption Dynamics for Stayers - Propensity Score}
        \label{tab:table_full_nmv_match}
\end{table}

\begin{table}[ht!]
    \centering
    \resizebox{12cm}{!}{
\begin{tabular}{lccccc} \hline
 & (1) & (2) & (3) & (4) & (5) \\
VARIABLES & Log Consumption & Log Mortgage Bal. & Mortgage Orig. & Log Auto Bal. & Auto Orig. \\ \hline
\vspace{4pt} & \begin{footnotesize}\end{footnotesize} & \begin{footnotesize}\end{footnotesize} & \begin{footnotesize}\end{footnotesize} & \begin{footnotesize}\end{footnotesize} & \begin{footnotesize}\end{footnotesize} \\
3 yrs until shock & 0.00509 & 0.0832** & 0.00629** & 0.101 & 0.00826 \\
\vspace{4pt} & \begin{footnotesize}(0.0367)\end{footnotesize} & \begin{footnotesize}(0.0422)\end{footnotesize} & \begin{footnotesize}(0.00251)\end{footnotesize} & \begin{footnotesize}(0.0911)\end{footnotesize} & \begin{footnotesize}(0.0114)\end{footnotesize} \\
2 yrs until shock & 0.0424 & 0.0387 & 0.00141 & 0.0153 & -0.00623 \\
\vspace{4pt} & \begin{footnotesize}(0.0289)\end{footnotesize} & \begin{footnotesize}(0.0317)\end{footnotesize} & \begin{footnotesize}(0.00132)\end{footnotesize} & \begin{footnotesize}(0.0735)\end{footnotesize} & \begin{footnotesize}(0.00855)\end{footnotesize} \\
shock & 0.105*** & 1.507*** & 0.121*** & 0.127 & 0.0190** \\
\end{tabular}
}
        \captionof{table}{HPI Shock on Consumption Dynamics for Movers  ($t=0$)  - Propensity Score}
        \label{tab:shock_move_match}
\end{table}

\clearpage
\newpage

\newpage
\section{DID} 
\label{sec:DID}

We start off with the simplest specification, which is described below in Equation \ref{eq:DID}
\begin{equation}\label{eq:DID}
y_{ict}=\alpha+\beta HPI_{ct}+ b X_{ict} +\mu_i+\epsilon_{ict}
\end{equation}

\newpage

    \begin{table}[ht!]
  \centering
  \resizebox{12cm}{!}{
\begin{tabular}{lccccc} \hline
 & (1) & (2) & (3) & (4) & (5) \\
VARIABLES & Log Consumption & Log Mortgage Bal. & Mortgage Orig. & Log Auto Bal. & Auto Orig. \\ \hline
\vspace{4pt} & \begin{footnotesize}\end{footnotesize} & \begin{footnotesize}\end{footnotesize} & \begin{footnotesize}\end{footnotesize} & \begin{footnotesize}\end{footnotesize} & \begin{footnotesize}\end{footnotesize} \\
Post & 0.0354*** & -0.0500*** & -0.000961** & 0.0224 & -0.000875 \\
 & \begin{footnotesize}(0.00595)\end{footnotesize} & \begin{footnotesize}(0.0107)\end{footnotesize} & \begin{footnotesize}(0.000410)\end{footnotesize} & \begin{footnotesize}(0.0168)\end{footnotesize} & \begin{footnotesize}(0.00116)\end{footnotesize} \\
\vspace{4pt} & \begin{footnotesize}\end{footnotesize} & \begin{footnotesize}\end{footnotesize} & \begin{footnotesize}\end{footnotesize} & \begin{footnotesize}\end{footnotesize} & \begin{footnotesize}\end{footnotesize} \\
Observations & 1,372,704 & 1,617,767 & 1,483,291 & 1,617,767 & 1,483,291 \\
Shock & HPI & HPI & HPI & HPI & HPI \\
No. ID & 203272 & 219891 & 214125 & 219891 & 214125 \\
 Dep. Var Mean & 7.764 & .3133 & .01279 & 2.96 & .0825 \\ \hline
\multicolumn{6}{c}{\begin{footnotesize} Robust standard errors in parentheses\end{footnotesize}} \\
\multicolumn{6}{c}{\begin{footnotesize} *** p$<$0.01, ** p$<$0.05, * p$<$0.1\end{footnotesize}} \\
\end{tabular}
  }
  \captionof{table}{DID: HPI Shock on Consumption}
\label{tab:DID_HPI_Consumption}
   \end{table}

     \begin{table}[ht!]
  \centering
  \resizebox{12cm}{!}{
\begin{tabular}{lccccc} \hline
 & (1) & (2) & (3) & (4) & (5) \\
VARIABLES & Log Consumption & Log Mortgage Bal. & Mortgage Orig. & Log Auto Bal. & Auto Orig. \\ \hline
\vspace{4pt} & \begin{footnotesize}\end{footnotesize} & \begin{footnotesize}\end{footnotesize} & \begin{footnotesize}\end{footnotesize} & \begin{footnotesize}\end{footnotesize} & \begin{footnotesize}\end{footnotesize} \\
Post & 0.00930 & -0.561*** & -0.0203*** & -0.0263 & -0.00185 \\
 & \begin{footnotesize}(0.00731)\end{footnotesize} & \begin{footnotesize}(0.00664)\end{footnotesize} & \begin{footnotesize}(0.000236)\end{footnotesize} & \begin{footnotesize}(0.0203)\end{footnotesize} & \begin{footnotesize}(0.00135)\end{footnotesize} \\
\vspace{4pt} & \begin{footnotesize}\end{footnotesize} & \begin{footnotesize}\end{footnotesize} & \begin{footnotesize}\end{footnotesize} & \begin{footnotesize}\end{footnotesize} & \begin{footnotesize}\end{footnotesize} \\
Observations & 1,186,616 & 1,397,989 & 1,289,000 & 1,397,989 & 1,289,000 \\
Shock & HPI and no move & HPI and no move & HPI and no move & HPI and no move & HPI and no move \\
No. ID & 180216 & 195859 & 191055 & 195859 & 191055 \\
 Dep. Var Mean & 7.758 & .2343 & .009542 & 2.909 & .08061 \\ \hline
\multicolumn{6}{c}{\begin{footnotesize} Robust standard errors in parentheses\end{footnotesize}} \\
\multicolumn{6}{c}{\begin{footnotesize} *** p$<$0.01, ** p$<$0.05, * p$<$0.1\end{footnotesize}} \\
\end{tabular}
  }
  \captionof{table}{DID: HPI Shock on Consumption for Stayers}
\label{tab:HPI_Consumption_stayers}
   \end{table}

\begin{table}[ht!]
  \centering
  \resizebox{12cm}{!}{
\begin{tabular}{lccccc} \hline
 & (1) & (2) & (3) & (4) & (5) \\
VARIABLES & Log Consumption & Log Mortgage Bal. & Mortgage Orig. & Log Auto Bal. & Auto Orig. \\ \hline
\vspace{4pt} & \begin{footnotesize}\end{footnotesize} & \begin{footnotesize}\end{footnotesize} & \begin{footnotesize}\end{footnotesize} & \begin{footnotesize}\end{footnotesize} & \begin{footnotesize}\end{footnotesize} \\
Post & 0.124*** & 1.300*** & 0.0388*** & 0.197*** & 0.00392* \\
 & \begin{footnotesize}(0.0109)\end{footnotesize} & \begin{footnotesize}(0.0322)\end{footnotesize} & \begin{footnotesize}(0.00115)\end{footnotesize} & \begin{footnotesize}(0.0311)\end{footnotesize} & \begin{footnotesize}(0.00222)\end{footnotesize} \\
\vspace{4pt} & \begin{footnotesize}\end{footnotesize} & \begin{footnotesize}\end{footnotesize} & \begin{footnotesize}\end{footnotesize} & \begin{footnotesize}\end{footnotesize} & \begin{footnotesize}\end{footnotesize} \\
Observations & 1,015,439 & 1,189,482 & 1,100,845 & 1,189,482 & 1,100,845 \\
Shock & HPI and move & HPI and move & HPI and move & HPI and move & HPI and move \\
No. ID & 164462 & 178192 & 174070 & 178192 & 174070 \\
 Dep. Var Mean & 7.753 & .3641 & .01418 & 3.001 & .08391 \\ \hline
\multicolumn{6}{c}{\begin{footnotesize} Robust standard errors in parentheses\end{footnotesize}} \\
\multicolumn{6}{c}{\begin{footnotesize} *** p$<$0.01, ** p$<$0.05, * p$<$0.1\end{footnotesize}} \\
\end{tabular}
}
  \captionof{table}{DID: HPI Shock on Consumption for Movers}
\label{tab:HPI_Consumption_movers}
   \end{table}

\end{document}